\newcommand{\defeq}[1]{\begin{equation}\label{#1}}
\newcommand{\eeq}{\end{equation}}

\documentclass[twocolumn]{article}

\usepackage{amsfonts}
\usepackage{amsmath}

\usepackage{amsmath}
\usepackage{amssymb}
\usepackage{amsthm}

\usepackage{framed}
\usepackage{algorithmic}

\usepackage[margin=1.0in]{geometry}
\usepackage{multicol}
\usepackage{multirow}

\usepackage{etoolbox}

\usepackage{color} 
\definecolor{shadecolor}{gray}{0.9}
\definecolor{lightyellow}{RGB}{255,255,192}
\definecolor{lightblue}{RGB}{192,192,255}
\definecolor{lightpurple}{RGB}{255,192,255}
\definecolor{lightgreen}{RGB}{192,255,192}

\usepackage{float}
\usepackage{caption}[2007/09/01] 

\usepackage{tikz}
	\usetikzlibrary{arrows}
	\usetikzlibrary{snakes}
	\usetikzlibrary{decorations.fractals}

\numberwithin{equation}{section}	

\theoremstyle{plain}			

\theoremstyle{definition}		

\renewcommand{\qed}{\hfill \mbox{\raggedright \rule{.07in}{.1in}}}

\DeclareCaptionLabelFormat{underlcap}{\bf{$\underline{\text{Figure #2}}$}}
\newenvironment{ColumnFigure}
{
	\begin{table}
	\begin{framed}
}
{
	\end{framed}
	\end{table}
}
\newenvironment{PageFigure}
{
	\begin{table*}
	\begin{framed}
}
{
	\end{framed}
	\end{table*}
}
\newcommand\Section[2][]{   \ifx\relax#1\relax\section{#2}\else\section[#1]{#2}\fi }

\newcommand\Subsection[2][]{   \ifx\relax#1\relax\subsection{#2}\else\subsection[#1]{#2}\fi   }

\patchcmd{\thebibliography}{\section*}{\Section}{}{}

\newcommand{\TitleString}{Towards $P=NP$ via k-SAT} 
\title{\TitleString:\\A k-SAT Algorithm Using Linear Algebra on Finite Fields
\\
\copyright \small{Matt Groff 2011} \\
\textsc{Semi-complete First Draft} \
} 
\author{MATT GROFF\\
P.O. Box 642\\
Camp Hill, PA, USA 17001-0642\\
mgroff100@hotmail.com}

\begin{document}
\pagestyle{myheadings}
\markright{\rlap{\TitleString}\hfill \copyright \textnormal{Matt Groff 2011}\hfill}
\twocolumn[\maketitle]
\begin{framed}
\begin{abstract}
The problem of P vs. NP is very serious, and solutions to the problem can help save lives.  This article is an attempt at solving the problem using a computer algorithm.
It is presented in a fashion that will hopefully allow for easy understanding for many people and scientists from many diverse fields.

In technical terms,
a novel method for solving k-SAT is explained.
This method is primarily based on linear algebra and finite fields.
Evidence is given that this method may require rougly {\color{red} O($n^3$)} time and space for deterministic models.  More specifically the algorithm runs in time $O(P\cdot V(n+V)^2)$ with mistaking satisfiable Boolean expressions as unsatisfiable with an approximate probablity $1 / \Theta(V(n+V)^2)^P$, where $n$ is the number of clauses and $V$ is the number of variables.  It's concluded that significant evidence exists that P=NP.

There is a forum devoted to this paper at {\emph{http://482527.ForumRomanum.com}}.  All are invited to correspond here and help with the analysis of the algorithm.  Source code for the associated algorithm can be found at {\emph{https://sourceforge.net/p/la3sat}}.
\end{abstract}
\end{framed}

\Section{Introduction}
There are many problems proposed to computer scientists that have been thought to be too difficult for computers to solve quickly.  In fact, perhaps the most fundamental question in computer science is to find if certain types of problems, collectively known as the class NP, can be solved quickly by a computer.  If so, a world of opportunities would open up, and many new problems that were supposed to be almost impossible to solve could be solved quickly.  This paper attempts to provide a proof that they can be solved quickly, and also shows a way to do it.  This will hopefully invite researchers from many diverse fields to contribute to the research and work of solving NP hard problems.

The level of interest in this question is so great that in 2000, the Clay Mathematics Institute listed the 7 Millennium Prize Problems, and offered \$1,000,000 for someone who could prove the relationship between P and NP\cite{D03}, which would answer the question.

\Subsection{Turing Machines}

\begin{PageFigure}
A decision tree is shown below.  As an algorithm (computer program) progresses, it must pick a choice from three possibilities.  Nondeterministic turing machines(NTMs) have been thought to have an advantage in this case, because, at each step, they can pick from any choice.  Deterministic turing machines(DTMs) don't really have the ability to pick, so they're thought to be at a disadvantage.
	\newline
	\begin{center}
		\begin{tikzpicture}
		
			\usetikzlibrary{arrows}
			\tikzstyle{L1}=[circle,draw=red,fill=red!20,thick]
			\tikzstyle{L2best}=[circle,draw=red,fill=red!20,thick, inner sep=0pt,minimum size=25]]
			\tikzstyle{L2other}=[circle,draw=green,fill=green!20,thick,inner sep=0pt,minimum size=25]]
			\tikzstyle{L3best}=[circle,draw=red,fill=red!20,thick, inner sep=0pt,minimum size=15]]
			\tikzstyle{L3other}=[circle,draw=green,fill=green!20,thick,inner sep=0pt,minimum size=15]]
			\tikzstyle{L4best}=[circle,draw=red,fill=red!20,thick, inner sep=0pt,minimum size=10]]
			\tikzstyle{L4other}=[circle,draw=green,fill=green!20,thick,inner sep=0pt,minimum size=10]]
			\node (START)  at ( 0,0)  [L1]      {START};
			\node (L2left) at (-4.5,-2) [L2other] { };
			\node (L2mid) at ( 0,-2) [L2best] { };
			\node (L2right) at ( 4.5,-2) [L2other] { };
				\node (L3-6) at (-6,-3.5) [L3other] { };
				\node (L3-4p5) at (-4.5,-3.5) [L3other] { };
				\node (L3-3) at (-3,-3.5) [L3other] { };
			\node (L3-1p5) at (-1.5,-3.5) [L3other] { };
			\node (L30) at (  0,-3.5) [L3other] { };
			\node (L31p5) at (1.5,-3.5) [L3best] { };
				\node (L36) at (6,-3.5) [L3other] { };
				\node (L34p5) at (4.5,-3.5) [L3other] { };
				\node (L33) at (3,-3.5) [L3other] { };
				\node (L4-6p5) at (-6.5,-4.5) [L4other] { };
				\node (L4-6) at (-6,-4.5) [L4other] { };
				\node (L4-5p5) at (-5.5,-4.5) [L4other] { };
				\node (L4-5) at (-5,-4.5) [L4other] { };
				\node (L4-4p5) at (-4.5,-4.5) [L4other] { };
				\node (L4-4) at (-4,-4.5) [L4other] { };
				\node (L4-3p5) at (-3.5,-4.5) [L4other] { };
				\node (L4-3) at (-3,-4.5) [L4other] { };
				\node (L4-2p5) at (-2.5,-4.5) [L4other] { };
				\node (L4-2) at (-2,-4.5) [L4other] { };
				\node (L4-1p5) at (-1.5,-4.5) [L4other] { };
				\node (L4-1) at (-1,-4.5) [L4other] { };
				\node (L4-p5) at (-.5,-4.5) [L4other] { };
				\node (L40) at (0,-4.5) [L4other] { };
				\node (L4p5) at (.5,-4.5) [L4other] { };
			\node (L41) at (  1,-4.5) [L4best] { };
			\node (L41p5) at (1.5,-4.5) [L4other] { };
			\node (L42) at (  2,-4.5) [L4other] { };
			\node (L46p5) at (6.5,-4.5) [L4other] { };
				\node (L46) at (6,-4.5) [L4other] { };
				\node (L45p5) at (5.5,-4.5) [L4other] { };
				\node (L45) at (5,-4.5) [L4other] { };
				\node (L44p5) at (4.5,-4.5) [L4other] { };
				\node (L44) at (4,-4.5) [L4other] { };
				\node (L43p5) at (3.5,-4.5) [L4other] { };
				\node (L43) at (3,-4.5) [L4other] { };
				\node (L42p5) at (2.5,-4.5) [L4other] { };
				
			\draw [thick,-triangle 45] (START) -- (L2left);
			\draw [thick,-triangle 45,red] (START) -- (L2mid);
			\draw [thick,-triangle 45] (START) -- (L2right);
				\draw [thick,-triangle 45] (L2left) -- (L3-6);
					\draw [thick,-triangle 45] (L3-6) -- (L4-6p5);
					\draw [thick,-triangle 45] (L3-6) -- (L4-6);
					\draw [thick,-triangle 45] (L3-6) -- (L4-5p5);
				\draw [thick,-triangle 45] (L2left) -- (L3-4p5);
					\draw [thick,-triangle 45] (L3-4p5) -- (L4-5);
					\draw [thick,-triangle 45] (L3-4p5) -- (L4-4p5);
					\draw [thick,-triangle 45] (L3-4p5) -- (L4-4);
				\draw [thick,-triangle 45] (L2left) -- (L3-3);
					\draw [thick,-triangle 45] (L3-3) -- (L4-3p5);
					\draw [thick,-triangle 45] (L3-3) -- (L4-3);
					\draw [thick,-triangle 45] (L3-3) -- (L4-2p5);
				\draw [thick,-triangle 45] (L2mid) -- (L3-1p5);
					\draw [thick,-triangle 45] (L3-1p5) -- (L4-2);
					\draw [thick,-triangle 45] (L3-1p5) -- (L4-1p5);
					\draw [thick,-triangle 45] (L3-1p5) -- (L4-1);
				\draw [thick,-triangle 45] (L2mid) -- (L30);
					\draw [thick,-triangle 45] (L30) -- (L4-p5);
					\draw [thick,-triangle 45] (L30) -- (L40);
					\draw [thick,-triangle 45] (L30) -- (L4p5);
				\draw [red, thick,-triangle 45] (L2mid) -- (L31p5);
					\draw [thick,-triangle 45] (L31p5) -- (L42);
					\draw [thick,-triangle 45] (L31p5) -- (L41p5);
					\draw [red,thick,-triangle 45] (L31p5) -- (L41);
				\draw [thick,-triangle 45] (L2right) -- (L36);
					\draw [thick,-triangle 45] (L36) -- (L46p5);
					\draw [thick,-triangle 45] (L36) -- (L46);
					\draw [thick,-triangle 45] (L36) -- (L45p5);
				\draw [thick,-triangle 45] (L2right) -- (L34p5);
					\draw [thick,-triangle 45] (L34p5) -- (L45);
					\draw [thick,-triangle 45] (L34p5) -- (L44p5);
					\draw [thick,-triangle 45] (L34p5) -- (L44);
				\draw [thick,-triangle 45] (L2right) -- (L33);
					\draw [thick,-triangle 45] (L33) -- (L43p5);
					\draw [thick,-triangle 45] (L33) -- (L43);
					\draw [thick,-triangle 45] (L33) -- (L42p5);
					
					\node [circle,draw=green,fill=green!20,thick,label=right:Choice available to NTM] at (-2.5,-5.5){};
					\node [circle,draw=red,fill=red!20,thick,label=right:Choice available to NTM and DTM] at (-2.5,-6){};
%
		\end{tikzpicture}
	\end{center}
	\caption{\bf{A Scenario Involving Choices}\label{Choice}}
\end{PageFigure}

To understand the classes P and NP, first consider the basic notion of a Turing machine, which is a scientific definition of a computer.  It has one or more tapes that hold information.  At any time, the Turing machine uses the tape or tapes to decide what to do next.  The machine has a prescribed set of instructions to it to help it decide.

To develop the ideas behind Turing machines more, a distinction is made between what types of decisions can be made.  A deterministic Turing machine (DTM) has a predetermined decision for every type of situation it encounters; thus the term deterministic.  A nondeterministic Turing machine (NTM), on the other hand, can have more than one action it can do for any given situation.  In other words, it's actions aren't determined ahead of time.  Therefore, it's said that it's nondeterminstic.

Figure \ref{Choice} makes the difference more clear.  NTMs have been thought to be able to solve more problems than DTMs because of the availability to make more choices.  Here the DTM has only one choice available at each step, while the DTM has three choices available.  So this evidence points to DTMs as being more limited in potential than NTMs.  In fact, there is enough difference between these two computers that different classes of computation (computational power) were proposed for each.

\pagebreak

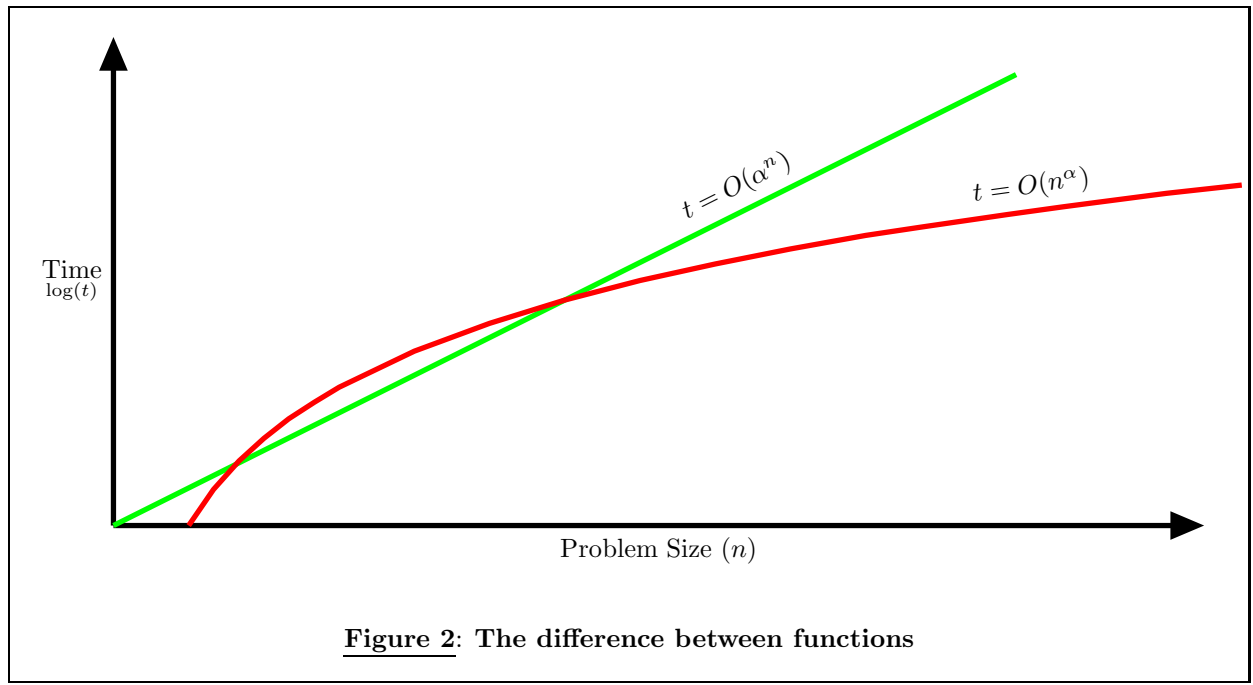
\Subsection{P and NP}
  
\begin{PageFigure}
\begin{center}
		\begin{tikzpicture}
			\draw[line width = 2, -triangle 45] (0,0) .. controls (0,3.25) .. node[left] {$\displaystyle\underset{\log(t)}{\mbox{Time}}$} (0,6.5);
			\draw[line width = 2, -triangle 45] (0,0) .. controls (7.25,0) .. node[below] {Problem Size ($n$)} (14.5,0);

			\draw [green, line width = 2] (0,0) ..controls (6,3)..node[sloped, above, near end, black]{$t=O(\alpha^n)$} (12,6);
			\draw [red, line width = 2] (1,0) -- (1.33,0.48) -- (1.67,0.86) -- (2,1.16) 
			  --(2.33,1.42) -- (2.67,1.64) -- (3,1.84) -- (4,2.32)
				-- (5,2.69) -- (6,3) -- (7,3.26) -- (8,3.48) -- (9,3.68) -- (10,3.86)
				.. controls (11, 4.01) and (12,4.16) .. 
				node[sloped, above, near end, black]{$t=O(n^\alpha)$}
				(13,4.29) -- (14, 4.42) -- (15,4.53);
		\end{tikzpicture}
	\end{center}
	\caption{\bf{The difference between functions}\label{Functions}}
\end{PageFigure}

Right around 1970, Leonid Levin, in the USSR, and Stephen Cook, of the US, independently arrived at the concept of NP-completeness\cite{NP-comp}.

The idea of NP-completeness comes from two classes of algorithms.  Once again, these two classes come from the ideas of DTMs and NTMs.

The classes are defined partially by how much can be accomplished within a limited amount of time.  This time limit, which will be explained briefly here, is defined by the mathematics of asymptotic analysis, which is described in \cite{Asymptote}.  Basically, the time limit is given as a function of the problem size.  The function can take on many forms, and two common forms are shown above in Figure \ref{Functions}.  Here a polynomial function ($t=O(n^\alpha)$) is contrasted with an exponential function ($t=O(\alpha^n)$).  Note that the polynomial functions may take more time for smaller problems, but the larger problems, which computer scientists are mainly concerned with, have smaller times for polynomial functions.

The two classes mentioned above are defined for polynomial functions.  One class, $P$, is defined as all problems that can be solved by a DTM in polynomial time (a polynomial function of the problem size).  The other class, $NP$, is defined as the class of all problems that can be solved by an NTM in polynomial time.  $NP$-complete problems are then problems in $NP$ that are at least as hard to solve as the hardest problems in $NP$.

The fundamental question that this paper attempts to answer is if $NP$-complete problems can be solved in polynomial time by a DTM; in other words, is $P=NP$?

For more information on $P$ versus $NP$, one can refer to Sipser \cite{PvsNP}.

\Subsection{SAT and k-SAT}
SAT essentially asks if, given a Boolean formula, can the formula be satisfied.  In other words, it asks if the variables in the formula be given a truth assignment that makes the entire formula true.  It's required, to thoroughly answer the question, that a certificate is given if the answer is true.  This is usually in the form of a particular solution to the question, such as a satisfying assignment for all variables.

k-SAT is  a particular variation of SAT, in which the formula is organized into clauses.  All variables inside a clause are connected via disjunction.  All clauses are connected via conjunction.  The $k$ in k-SAT then refers to the number of variables in each clause.  More information on SAT and k-SAT can be found in \cite{SATwiki}.

Conventional SAT and k-SAT solvers function by learning clauses or making random guesses at solutions (\cite{B09} and the recent survey \cite{PD10}).  

The current best upper bounds for 3-SAT is $O(1.32113^n)$ time, and can be found in ISAAC 2010 by Iwama et al\cite{3SATLower}.  There's an arXived paper by Hertli, Moser and Scheder which gives an $O(1.321^n)$ time algorithm\cite{3SATLower2}.  These are all randomized algorithms.  There is an arXived, deterministic 3-SAT algorithm by Kutzkov and Scheder that runs in time $O(1.439^n)$\cite{3SATLower3}.

There is a paper that proposed a polynomial runtime for a SAT variant.  V.F. Romanov proposed a 3-SAT algorithm that uses ``discordant structures'', or set-like operations on a lattice, to represent information and determine solutions in polynomial time\cite{Rom3SAT}.  However, Liao presents an argument that P is not equal to NP, using a 3-SAT variant, in \cite{Liao}.

Perhaps Baker, Gill, and Solovay's theory of relativization helps explain why there are not more algorithms that attempt to solve NP-complete problems in polynomial time\cite{BGS75}.  Their paper ``Relativizations of the P != NP Question'' states that consulting an oracle can lead to situations in which P != NP.

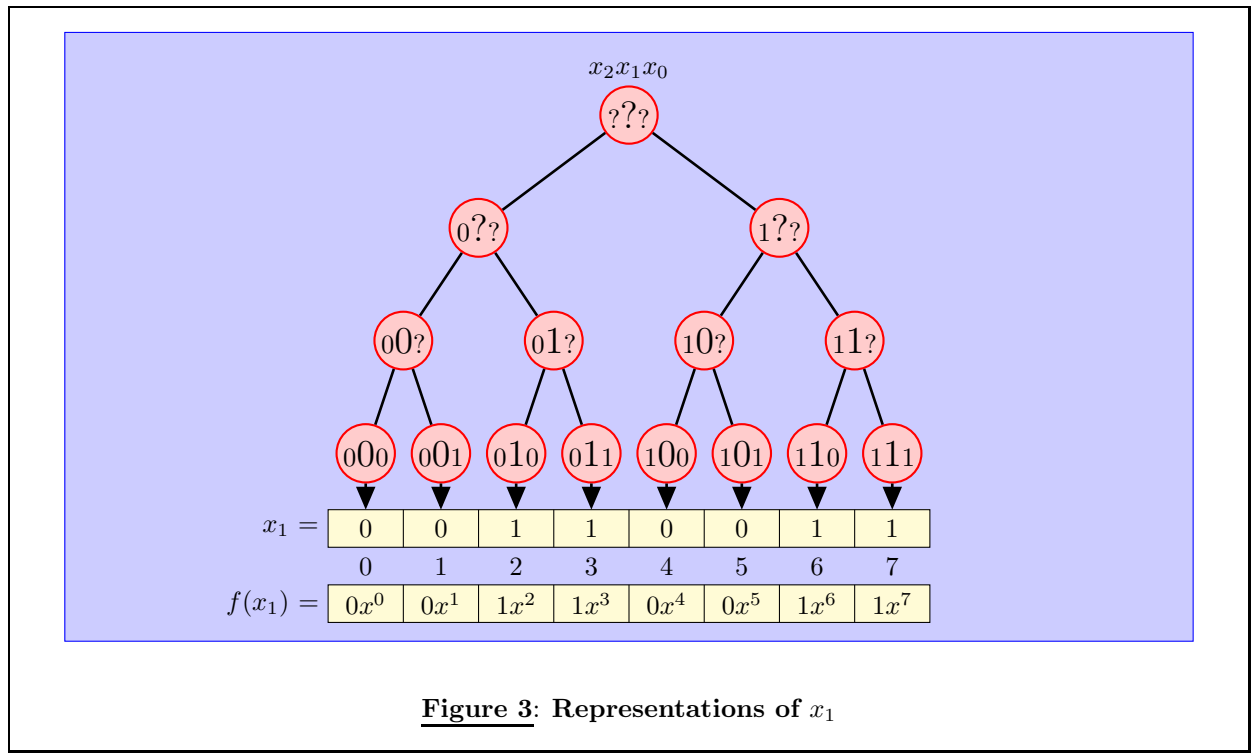
\begin{PageFigure}
	\begin{center}
		\begin{tikzpicture}
			\usetikzlibrary{arrows}
			
			\tikzstyle{N3}=[circle,draw=red,fill=red!20,thick, inner sep=1pt,minimum size=15];
			
			\filldraw [fill=blue!20, draw=blue] (-7.5,1.1) rectangle (7.5,-7);
			
			\node at (0,.6) {$x_2 x_1 x_0$};
			
			\node (xxx) at ( 0  ,   0) [N3] {?\Large{?}\normalsize?};
			\node (0xx) at (-2  ,-1.5) [N3] {0\Large{?}\normalsize?};
			\node (1xx) at ( 2  ,-1.5) [N3] {1\Large{?}\normalsize?};
			\node (00x) at (-3  ,-3  ) [N3] {0\Large{0}\normalsize?};
			\node (01x) at (-1  ,-3  ) [N3] {0\Large{1}\normalsize?};
			\node (10x) at ( 1  ,-3  ) [N3] {1\Large{0}\normalsize?};
			\node (11x) at ( 3  ,-3  ) [N3] {1\Large{1}\normalsize?};
			
			\node (000) at (-3.5,-4.5) [N3] {0\Large{0}\normalsize0};
			\node (001) at (-2.5,-4.5) [N3] {0\Large{0}\normalsize1};
			\node (010) at (-1.5,-4.5) [N3] {0\Large{1}\normalsize0};
			\node (011) at (-0.5,-4.5) [N3] {0\Large{1}\normalsize1};
			\node (100) at ( 0.5,-4.5) [N3] {1\Large{0}\normalsize0};
			\node (101) at ( 1.5,-4.5) [N3] {1\Large{0}\normalsize1};
			\node (110) at ( 2.5,-4.5) [N3] {1\Large{1}\normalsize0};
			\node (111) at ( 3.5,-4.5) [N3] {1\Large{1}\normalsize1};
			
			\draw [line width = 1] (xxx) -- (0xx);
			\draw [line width = 1] (xxx) -- (1xx);
			\draw [line width = 1] (0xx) -- (00x);
			\draw [line width = 1] (0xx) -- (01x);
			\draw [line width = 1] (1xx) -- (10x);
			\draw [line width = 1] (1xx) -- (11x);
			
			\draw [line width = 1] (00x) -- (000);
			\draw [line width = 1] (00x) -- (001);
			\draw [line width = 1] (01x) -- (010);
			\draw [line width = 1] (01x) -- (011);
			\draw [line width = 1] (10x) -- (100);
			\draw [line width = 1] (10x) -- (101);
			\draw [line width = 1] (11x) -- (110);
			\draw [line width = 1] (11x) -- (111);
			
			\filldraw [fill = yellow!20] (-4,-5.25) rectangle (4,-5.75);
			
			\foreach \x in {-3,-2,...,3}
				\draw (\x,-5.25) -- (\x,-5.75);
			
			\draw [line width = 1,-triangle 45] (000) -- (-3.5, -5.25);
			\draw [line width = 1,-triangle 45] (001) -- (-2.5, -5.25);
			\draw [line width = 1,-triangle 45] (010) -- (-1.5, -5.25);
			\draw [line width = 1,-triangle 45] (011) -- (-0.5, -5.25);
			\draw [line width = 1,-triangle 45] (100) -- ( 0.5, -5.25);
			\draw [line width = 1,-triangle 45] (101) -- ( 1.5, -5.25);
			\draw [line width = 1,-triangle 45] (110) -- ( 2.5, -5.25);
			\draw [line width = 1,-triangle 45] (111) -- ( 3.5, -5.25);
			
			
			\node at (-4,-5.5) [left] {$x_1$ =};
			
			\foreach \x in {0,1,...,7}
				\node  at (\x-3.5,-6) {\x};
				
			\node at (-3.5,-5.5) {0};
			\node at (-2.5,-5.5) {0};
			\node at (-1.5,-5.5) {1};
			\node at (-0.5,-5.5) {1};
			\node at ( 0.5,-5.5) {0};
			\node at ( 1.5,-5.5) {0};
			\node at ( 2.5,-5.5) {1};
			\node at ( 3.5,-5.5) {1};
			
			\filldraw [fill = yellow!20] (-4,-6.25) rectangle (4,-6.75);
			\foreach \x in {-3,-2,...,3}
				\draw (\x,-6.25) -- (\x,-6.75);
			\node at (-4,-6.5) [left] {$f(x_1)$ =};
			
			\node at (-3.5,-6.5) {$0x^0$};
			\node at (-2.5,-6.5) {$0x^1$};
			\node at (-1.5,-6.5) {$1x^2$};
			\node at (-0.5,-6.5) {$1x^3$};
			\node at ( 0.5,-6.5) {$0x^4$};
			\node at ( 1.5,-6.5) {$0x^5$};
			\node at ( 2.5,-6.5) {$1x^6$};
			\node at ( 3.5,-6.5) {$1x^7$};
			
			
		\end{tikzpicture}
	\end{center}
	\caption{\bf{Representations of $x_1$}\label{Representations}}
\end{PageFigure}

\pagebreak

\Section{Data Structure(s)}
One of the most fundamental components of the algorithm will be refered to as the \textit{clause polynomial}.  Figure \ref{Representations} shows the basic organization of clause polynomials.  Essentially, the complete information of one clause is represented by a polynomial of one variable.  That is, for eaxh particular truth assignment of all variables, the polynomial ``records''  whether the clause satisfies this assignment.  This is then attached to the polynomial's variable, which orders the truth assignments.

Note that the particular truth assingnments are shown in red.  In this fashion, the same tree that organizes the variables can be repeatedly used, and therefore provides some standard organization to the information.  Again, this tree organizes all possible variable assignements.  It does so by starting at the root, and then proceeds through all Boolean variables in order, and assigns one node to each possible assignment of true or false for each variable.  The leaves (at the bottom) thus represent a complete truth assingment for all variables.

\begin{ColumnFigure}
	\begin{center}
		\begin{tikzpicture}
			\filldraw [fill=blue!20, draw=blue] (-0.4,1.5) rectangle (5.6,-4);
			\node at (0.4, 1  ) {\Large{1 Bit}};
			\node at (0  , 0  ) [right] {\texttt{0}\textnormal(0)};
			\node at (0  ,-0.5) [right] {\texttt{1}\textnormal(1)};
			
			\node at (2.5, 1  ) {\Large{2 Bits}};
			\node at (2  , 0  ) [right] {\texttt{00}\textnormal(0)};
			\node at (2  ,-0.5) [right] {\texttt{01}\textnormal(1)};
			\node at (2  ,-1  ) [right] {\texttt{10}\textnormal(2)};
			\node at (2  ,-1.5) [right] {\texttt{11}\textnormal(3)};
			
			\node at (4.6, 1  ) {\Large{3 Bits}};
			\node at (4  , 0  ) [right] {\texttt{0\Large{0}\normalsize0}\textnormal(0)};
			\node at (4  ,-0.5) [right] {\texttt{0\Large{0}\normalsize1}\textnormal(1)};
			\node at (4  ,-1  ) [right] {\texttt{0\Large{1}\normalsize0}\textnormal(2)};
			\node at (4  ,-1.5) [right] {\texttt{0\Large{1}\normalsize1}\textnormal(3)};
			\node at (4  ,-2  ) [right] {\texttt{1\Large{0}\normalsize0}\textnormal(4)};
			\node at (4  ,-2.5) [right] {\texttt{1\Large{0}\normalsize1}\textnormal(5)};
			\node at (4  ,-3  ) [right] {\texttt{1\Large{1}\normalsize0}\textnormal(6)};
			\node at (4  ,-3.5) [right] {\texttt{1\Large{1}\normalsize1}\textnormal(7)};

		\end{tikzpicture}
	\end{center}
	\caption{\bf{Binary Representations}\label{BinaryRepresentations}}
\end{ColumnFigure}

\begin{PageFigure}
	\begin{center}
		\begin{tikzpicture}
			\filldraw [fill=blue!20,draw=blue] (-7.5,1) rectangle (7.5,-2);
			\node at (0, 0.5) {$x_0$ In System Of $x_0$, $x_1$, $x_2$};
			\draw (-2.1, 0.25) -- (2.1, 0.25);
			\node at (0,-0.5) {$x^{2^0}(1 + x^{2^1})(1 + x^{2^2}) =$};
			\node at (0,-1  ) {$x^1 + x^3 + x^5 + x^7 =$};
			\node at (0,-1.5) {$\mbox{\Large0}x^0 + \mbox{\Large1}x^1 + \mbox{\Large0}x^2 +
				\mbox{\Large1}x^3 + \mbox{\Large0}x^4 + \mbox{\Large1}x^5 +
				\mbox{\Large0}x^6 + \mbox{\Large1}x^7$};
				
			\filldraw [fill=purple!20,draw=purple] (-7.5,-2.5) rectangle (7.5,-5.5);
			\node at (0,-3  ) {$x_1$ In System Of $x_0$, $x_1$, $x_2$};
			\draw (-2.1,-3.25) -- (2.1, -3.25);
			\node at (0,-4.0) {$(1 + x^{2^0})x^{2^1}(1 + x^{2^2}) =$};
			\node at (0,-4.5) {$x^2 + x^3 + x^6 + x^7 =$};
			\node at (0,-5.0) {$\mbox{\Large0}x^0 + \mbox{\Large0}x^1 + \mbox{\Large1}x^2 +
				\mbox{\Large1}x^3 + \mbox{\Large0}x^4 + \mbox{\Large0}x^5 + 
				\mbox{\Large1}x^6 + \mbox{\Large1}x^7$}; 
				
		\end{tikzpicture}
	\end{center}
	\caption{\bf{Constructing Polynomials For Variables}\label{VariableConstruction}}
\end{PageFigure}

The clause polynomial for a single variable is equivalent to a digit in a binary number.  In order to understand this, Figure \ref{BinaryRepresentations} shows binary numbers composed of one to three bits.  The tree in Figure \ref{Representations} uses a three bit example, since there are three variables; thus one bit for each variable.  Note then, that the middle variable, $x_1$, out of $x_0$, $x_1$, and $x_2$, corresponds with the middle bit.  This can be seen in the figure by looking at the middle bit for each binary representation of three bits (which is displayed larger than the other two bits).  The sequence is the same as the binary tree above.

\begin{PageFigure}
	\begin{center}
		\begin{tikzpicture}
			
			\filldraw[fill=blue!20,draw=blue] (-0.65,1) rectangle (9.0,32*-.35);
			\filldraw[fill=purple!20,draw=purple] (9.1,1) rectangle (14.35,32*-.35);
			
			\node at (0  ,.5) {$x_4$};
			\node at (0.5,.5) {$x_3$};
			\node at (1  ,.5) {$x_2$};
			\node at (1.5,.5) {$x_1$};
			\node at (2  ,.5) {$x_0$};
			
			\node at (3.5,.5) {$x_1 \lor x_3$};
			\node at (5.0,.5) {$x_0 \lor x_3$};
			\node at (6.5,.5) {$x_0 \lor x_2$};
			\node at (8.0,.5) {$x_0 \lor x_1$};
			
			\node at (10.5,.5) {$x_0 \lor x_1 \lor x_3$};
			\node at (13  ,.5) {$x_0 \lor x_1 \lor x_2$};
			
			\draw (-0.5,.25) -- (14.2,.25);
			
			\filldraw [fill=yellow!20!blue!20] (3.0,0.2) rectangle (4.0,-0.2-15*.35);
			\draw [dashed] (3.0,-0.2-7*.35) -- (4.0,-0.2-7*.35);
			\filldraw [fill=yellow!20!blue!20] (4.5,0.2) rectangle (5.5,-0.2-15*.35);
			\draw [dashed] (4.5,-0.2-7*.35) -- (5.5,-0.2-7*.35);
			\filldraw [fill=yellow!20!blue!20] (6.0,0.2) rectangle (7.0,-0.2-7*.35);
			\draw [dashed] (6.0,-0.2-3*.35) -- (7.0,-0.2-3*.35);
			\filldraw [fill=yellow!20!blue!20] (7.5,0.2) rectangle (8.5,-0.2-3*.35);
			\draw [dashed] (7.5,-0.2-1*.35) -- (8.5,-0.2-1*.35);
			
			\filldraw [fill=yellow!20!purple!20] ( 9.75,0.22) rectangle (11.25,-0.2-15*.35);
			\draw [dashed] (9.75,-0.2-7*.35) -- (11.25,-0.2-7*.35);
			\filldraw [fill=yellow!20!purple!20] (10.0,0.2) rectangle (11.0,-0.2-3*.35);
			\draw [dashed] (10.0,-0.2-1*.35) -- (11.0,-0.2-1*.35);
			
			\filldraw [fill=yellow!20!purple!20] (12.25,0.22) rectangle (13.75,-0.2-7*.35);
			\filldraw [fill=yellow!20!purple!20] (12.5,0.2) rectangle (13.5,-0.2-3*.35);
			\draw [dashed] (12.5,-0.2-1*.35) -- (13.5,-0.2-1*.35);

			\draw (6.0,-0.2-15*.35) -- (7.0,-0.2-15*.35);
			\draw (6.0,-0.2-23*.35) -- (7.0,-0.2-23*.35);
			\draw (7.5,-0.2- 7*.35) -- (8.5,-0.2- 7*.35);
			\draw (7.5,-0.2-11*.35) -- (8.5,-0.2-11*.35);
			\draw (7.5,-0.2-15*.35) -- (8.5,-0.2-15*.35);
			\draw (7.5,-0.2-19*.35) -- (8.5,-0.2-19*.35);
			\draw (7.5,-0.2-23*.35) -- (8.5,-0.2-23*.35);
			\draw (7.5,-0.2-27*.35) -- (8.5,-0.2-27*.35);
			
			\draw (12.25,-0.2-15*.35) -- (13.75,-0.2-15*.35);
			\draw (12.25,-0.2-23*.35) -- (13.75,-0.2-23*.35);
			
			\foreach \y in {0,1,...,15}
				\node at (0,\y*-0.35) {\texttt0};
			\foreach \y in {16,17,...,31}
				\node at (0,\y*-0.35) {\texttt1};
			
			\foreach \y in {0,1,...,7,16,17,...,23}
				\node at (0.5,\y*-0.35) {\texttt0};
			\foreach \y in {8,9,...,15,24,25,...,31}
				\node at (0.5,\y*-0.35) {\texttt1};
			
			\foreach \y in {0,1,...,3,8,9,...,11,16,17,...,19,24,25,...,27}
				\node at (1.0,\y*-0.35) {\texttt0};
			\foreach \y in {4,5,...,7,12,13,...,15,20,21,...,23,28,29,...,31}
				\node at (1.0,\y*-0.35) {\texttt1};
				
			\foreach \y in {0,1,4,5,8,9,12,13,16,17,20,21,24,25,28,29}
				\node at (1.5,\y*-0.35) {\texttt0};
			\foreach \y in {2,3,6,7,10,11,14,15,18,19,22,23,26,27,30,31}
				\node at (1.5,\y*-0.35) {\texttt1};
				
			\foreach \y in {0,2,...,30}
				\node at (2.0,\y*-0.35) {\texttt0};	
			\foreach \y in {1,3,...,31}
				\node at (2.0,\y*-0.35) {\texttt1};
				
%

			\foreach \y in {0,1,4,5}
				\node at (3.5,\y*-0.35) {\texttt0};
			\foreach \y in {2,3,6,7,...,15}
				\node at (3.5,\y*-0.35) {\texttt1};
				
			{
				\foreach \y in {0,1,4,5}
					\node at (3.5,\y*-0.35+16*-0.35) {\texttt0};
				\foreach \y in {2,3,6,7,...,15}
					\node at (3.5,\y*-0.35+16*-0.35) {\texttt1};
			}

			\foreach \y in {0,2,...,6}
				\node at (5.0,\y*-0.35) {\texttt0};
			\foreach \y in {1,3,...,7}
				\node at (5.0,\y*-0.35) {\texttt1};
			\foreach \y in {8,10,...,14}
				\node at (5.0,\y*-0.35) {\texttt1};
			\foreach \y in {9,11,...,15}
				\node at (5.0,\y*-0.35) {\texttt1};
			
			{
				\foreach \y in {0,2,...,6}
					\node at (5.0,\y*-0.35+16*-0.35) {\texttt0};
				\foreach \y in {1,3,...,7}
					\node at (5.0,\y*-0.35+16*-0.35) {\texttt1};
				\foreach \y in {8,10,...,14}
					\node at (5.0,\y*-0.35+16*-0.35) {\texttt1};
				\foreach \y in {9,11,...,15}
					\node at (5.0,\y*-0.35+16*-0.35) {\texttt1};
			}

			\foreach \z in {0,16}
			{
				\foreach \y in {0,2}
					\node at (6.5,\y*-0.35+\z*-0.35) {\texttt0};
				\foreach \y in {1,3,4,...,7}
					\node at (6.5,\y*-0.35+\z*-0.35) {\texttt1};
			}
			\foreach \z in {8,24}
			{
				\foreach \y in {0,2}
					\node at (6.5,\y*-0.35+\z*-0.35) {\texttt0};
				\foreach \y in {1,3,4,...,7}
					\node at (6.5,\y*-0.35+\z*-0.35) {\texttt1};
			}
			
			\foreach \z in {0,8,16,24}
			{
				\node at (8.0,\z*-0.35) {\texttt0};
				\foreach \y in {1,2,3}
					\node at (8.0,\y*-0.35+\z*-0.35) {\texttt1};
			}
			\foreach \z in {4,12,20,28}
			{
				\node at (8.0,\z*-0.35) {\texttt0};
				\foreach \y in {1,2,3}
					\node at (8.0,\y*-0.35+\z*-0.35) {\texttt1};
			}
			
			\foreach \y in {0,4,16,20}
				\node at (10.5,\y*-0.35) {\texttt0};
			\foreach \y in {1,2,3,5,6,...,15,17,18,19,21,22,...,31}
				\node at (10.5,\y*-0.35) {\texttt1};
				
			\foreach \y in {0,8,16,24}
				\node at (13,\y*-0.35) {\texttt0};
			\foreach \y in {1,2,...,7,9,10,...,15,17,18,...,23,25,26,...,31}
				\node at (13,\y*-0.35) {\texttt1};

		\end{tikzpicture}
	\end{center}
	\caption{\bf{Repeating Patterns of Clauses}\label{ClausePatterns}}
\end{PageFigure}

\begin{PageFigure}
	\begin{center}
		\begin{tikzpicture}
			\filldraw [fill=blue!20, draw=blue] (-3.5,1.5) rectangle (11.5,-4.5);
			\node at (4,1) {\Large{For Clauses of the Form $x_{a_0} \lor x_{a_1} \lor \dots \lor x_{a_z}$}};
			\draw (-1,0.75) -- (9, 0.75);
			\node at (0, 0  ) [right] {\textsc{For each $x_{a_i}$}};
			\node at (1,-0.5) [right] {\textsc{ Calculate
				$g(x_{a_i})=\left(\prod_{k=0}^{a_i-1}(1+x^{2^k})\right)x^{2^{a_i}}$}};
			\node at (0,-1.0) [right] {\textsc{Let} Result = 0};
			\node at (0,-1.5) [right] {\textsc{For $h=0$ to $z$}};
			\node at (1,-2.0) [right] {\textsc{If $(h = a_i)$ for some $a_i$}};
			\node at (2,-2.5) [right] {Result = Result + $g(x_h)$};
			\node at (1,-3.0) [right] {\textsc{Else}};
			\node at (2,-3.5) [right] {Result = Result$ \cdot (1 + x^{2^h})$};
			\node at (0,-4.0) [right] {\textsc{Return} Result};
			
		\end{tikzpicture}
	\end{center}
	\caption{\bf{Clause Polynomial Construction Algorithm}\label{ClauseAlgorithm}}
\end{PageFigure}

\begin{PageFigure}
	\begin{center}
		\begin{tikzpicture}
			\filldraw[fill=blue!20,draw=blue] (-2.5,1.5) rectangle (12.5,-2);
			\node at (5.0, 1  ) {\large{$(x_0 \lor x_1)$ From $x_0, x_1, x_2$}};
			\draw (2.8,.70) -- (7.2,.70);
			\node at (0  , 0  ) [right] {$g(x_0) = x^{2^0} = 1x^1$};
			\node at (4.5, 0  ) [right] {$g(x_1) = \left(1 + x^{2^0}\right)x^{2^1} = 1x^2 + 1x^3$};
			\node at (0  ,-0.75) [right] {$f(x_0 \lor x_1) = \left((x^1)+(x^2+x^3)\right)
				\left(1+x^{2^2}\right)$};
			\node at (0  ,-1.50) [right] {$f(x_0 \lor x_1) = {\mbox{\Large0}}x^0 + {\mbox{\Large1}}x^1 +
				{\mbox{\Large1}}x^2 + {\mbox{\Large1}}x^3 + {\mbox{\Large0}}x^4 + {\mbox{\Large1}}x^5 +
				{\mbox{\Large1}}x^6 + {\mbox{\Large1}}x^7$};
			
			\filldraw[fill=purple!20,draw=purple] (-2.5,-2.25) rectangle (12.5,-6.25);
			\node at (4.95,-2.75) {\large{$(x_0 \lor x_1 \lor x_2)$ From $x_0, x_1, x_2$}};
			\draw (2.3,-3.05) -- (7.7,-3.05);
			\node at (0  ,-3.75) [right] {$g(x_0) = x^{2^0} = 1x^1$};
			\node at (4.5,-3.75) [right] {$g(x_1) = \left(1 + x^{2^0}\right)x^{2^1} = 1x^2 + 1x^3$};
			\node at (0  ,-4.25) [right] {$g(x_2) = \left(1 + x^{2^0}\right)\left(1 + x^{2^1}\right)
				x^{2^2} = 1x^4 + 1x^5 + 1x^6 + 1x^7$};
			\node at (0  ,-5.00) [right] {$f(x_0 \lor x_1 \lor x_2) = (x^1) + (x^2 + x^3) +
				(x^4 + x^5 + x^6 + x^7)$};
			\node at (0  ,-5.75) [right] {$f(x_0 \lor x_1 \lor x_2) = {\mbox{\Large0}}x^0 +
				{\mbox{\Large1}}x^1 + {\mbox{\Large1}}x^2 + {\mbox{\Large1}}x^3 + {\mbox{\Large1}}x^4 +
				{\mbox{\Large1}}x^5 + {\mbox{\Large1}}x^6 + {\mbox{\Large1}}x^7$};
		\end{tikzpicture}
	\end{center}
	\caption{\bf{Clause Polynomial Examples}\label{ClauseExamples}}
\end{PageFigure}

\Subsection{Constructing Clause \\ Polynomials}
A basic clause is of the form

\defeq{ClauseEquation}
x_{a_0} \lor x_{a_1} \lor \dots \lor x_{a_z}
\eeq
The algorithm seeks to construct clause polynomials for clauses in this form.  This is a two step process.  The basic theory behind the process is fairly easily explained here, although the technicalities and a proof are saved for Appendix \ref{FormulaAppendix} on page \pageref{FormulaAppendix}.

To begin to understand how clause polynomials are constructed, a few patterns are observed.  Figure \ref{ClausePatterns}, on the next page, helps to demonstrate these.  First, the obvious pattern of binary digits appears for the variables on the lefthand side.  All variables have a pattern of $2^i$ zeros and then $2^i$ ones for variable $x_i$, which then repeats.  These variables, in conjunction, correspond with binary numbers.  In other words, for the least significant bit, the pattern is zero, then one, and then it repeats.  For the next least significant bit (or variable), the pattern has two zeros and then two ones, which then repeats.  This pattern continues for all variables.

Observing this pattern, an easy way to construct the clause polynomials for a single variable becomes clear.  In fact, it can be summarized as a simple formula:

\defeq{XmEquation}
f(x_m)=\left(\prod_{k=0}^{m-1}(1+x^{2^k})\right)x^{2^m}\left(\prod_{k=m+1}^{n}(1+x^{2^k})\right)
\eeq

Here $\prod{(f(x))}$ denotes the (indefinite) product of a function of x.  More information can be found about indefinite products in \cite{ProductWiki}.  This is taken over a system of $n+1$ variables.  

Figure \ref{VariableConstruction} shows how the clause polynomials are actually constructed using the formula.  Note that these are all for a single variable.  For the $x_0$ example, $m=0$ since the variable is $x_0$.  In other words, if the variable was $x_{32}$, then the algorithm would plug in $m=32$ into the formula.  Note, then, that $n$ is one less than the total number of variables, so in both cases it is two, since there are three variables.  Returning to the $x_0$ example, the lefthand product has nothing inside it, the middle follows from knowing $n$, and the right side product is determined from $m$ and $n$ and the rules of products.  For the $x_1$ example, there is a product on the left side to work with, since $m$ is now one, and $k$ is set to zero for this.  Then it follows that $(1+x^{2^k}) = (1+x^1)$ since $k=0$.  The right side follows similarly, this time with $k=2$.

Returning attention to Figure \ref{ClausePatterns} on the previous page, a second pattern can be observed.  This is appearant in the blue portion of the figure.  Here, the second pattern is shown in boxes for two-variable clauses.  There is the original one-variable pattern of ones and zeros, which is then followed by a series of ones.  Note that the series of ones is exactly the same size as the original single-variable pattern.  This entire pattern then repeats.

The observed pattern comes from the combination of two variables.  The long series of ones comes from the variable that is ``larger'' than the other.  The short series of repeating ones and zeros comes from the ``smaller'' of the two variables.  Together, they form a series that repeats.  Another way of looking at it is that both variables form a repeating series, and combining these two repeating series creates another repeating series.

In fact, the pattern of repetition continues even as more variables are added to the clause.  The purple portion of the figure shows the repitition(s) involved with three variables.  Again, combining two variables produces a short, repeating series.  When a third variable (that is strictly ``larger'' than the other two) is added, a longer, yet repeating, series emerges.

Again, the techinicalities of all of this are presented (and solved) in Appendix \ref{FormulaAppendix} on page \pageref{FormulaAppendix}.

There is a very simple algorithm to calculate clause polynomials for any amount of variables, as long as they are in the form given in Equation \ref{ClauseEquation}.  Figure \ref{ClauseAlgorithm} shows the algorithm.  The basic idea is to first calculate the the (possibly long) sequence of ones that repeats for each variable.  This is then shifted into the proper position.  It is identical to the clause calculations for one variable, with the exception that the series is not made to repeat.  The idea is that using ``smaller'' variables, the algorithm will construct short repeating sequences, and add in the appropriate larger variable when the repeating sequence gets large enough.  So essentially, it constructs one large repeating sequence by making the smaller variable sequences repeat, and adding in larger variables when the sequence gets large enough.

Figure \ref{ClauseExamples} shows examples of creating clause polynomials for given problems.  Note that the complete set of variables for the original problem must be known ahead of time, just as for clauses of individual variables.  As can be seen, the results from this figure correspond with the first eight values from the corresponding clauses in Figure \ref{ClausePatterns}.


There is really only one operation remaining to being able to construct essentially any clause polynomial of the form in Equation \ref{ClauseEquation}; negation.  As it turns out, negation is not much more difficult than constructing non-negated clause polynomials.

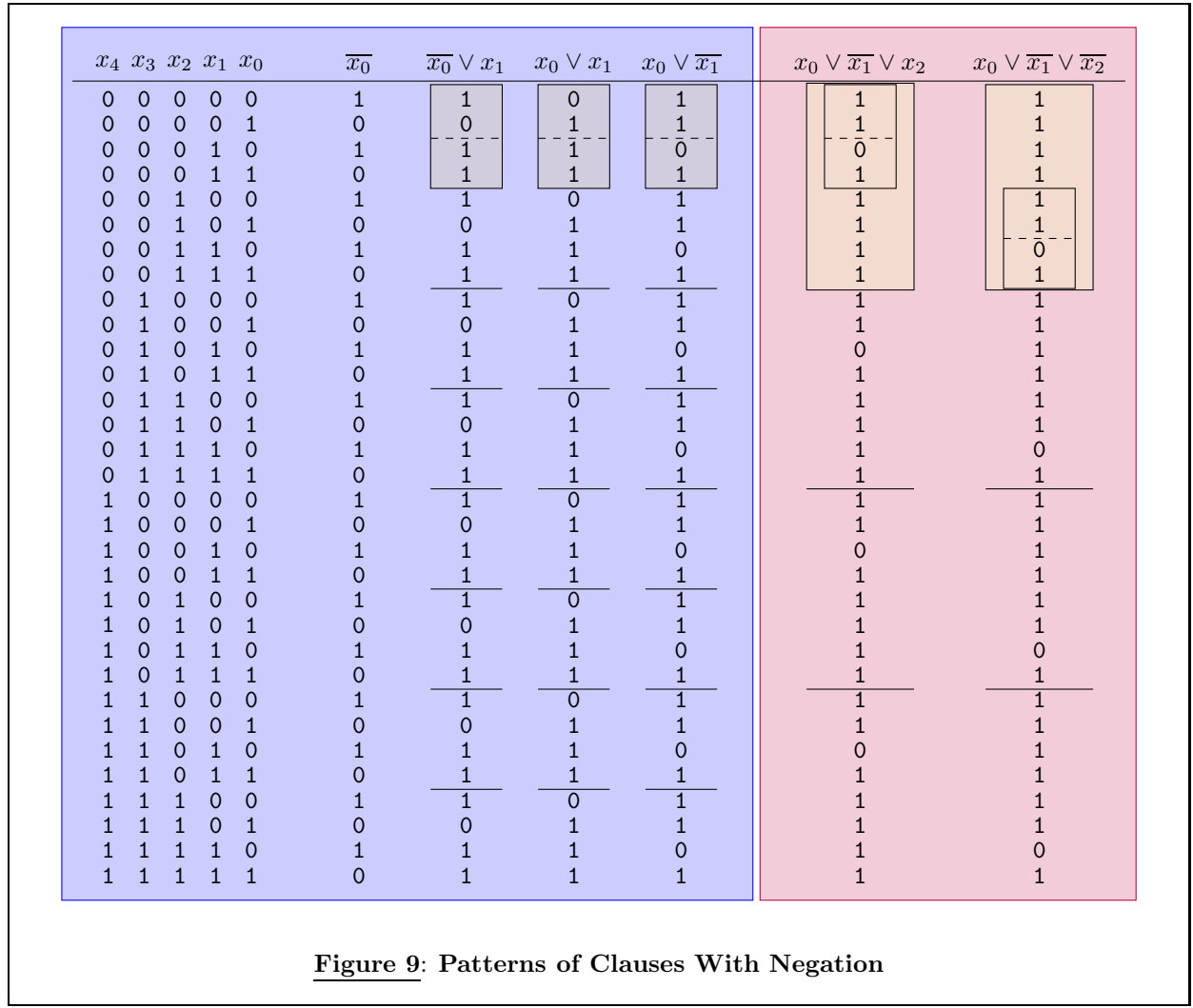
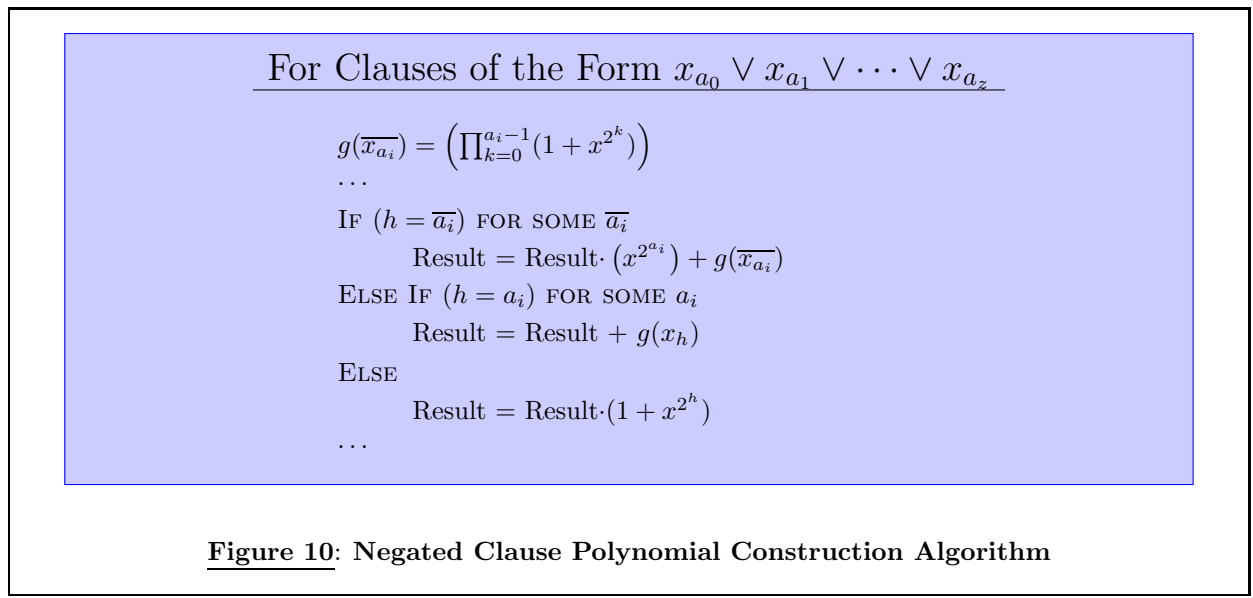
\begin{PageFigure}
	\begin{center}
		\begin{tikzpicture}
			
			\filldraw[fill=blue!20,draw=blue] (-0.65,1) rectangle (9.0,32*-.35);
			\filldraw[fill=purple!20,draw=purple] (9.1,1) rectangle (14.35,32*-.35);
			
			\node at (0  ,.5) {$x_4$};
			\node at (0.5,.5) {$x_3$};
			\node at (1  ,.5) {$x_2$};
			\node at (1.5,.5) {$x_1$};
			\node at (2  ,.5) {$x_0$};
			
			\node at (3.5,.5) {$\overline{x_0}$};
			\node at (5.0,.5) {$\overline{x_0} \lor x_1$};
			\node at (6.5,.5) {$x_0 \lor x_1$};
			\node at (8.0,.5) {$x_0 \lor \overline{x_1}$};
			
			\node at (10.5,.5) {$x_0 \lor \overline{x_1} \lor x_2$};
			\node at (13  ,.5) {$x_0 \lor \overline{x_1} \lor \overline{x_2}$};
			
			\draw (-0.5,.25) -- (14.2,.25);
			
			\filldraw [fill=yellow!20!blue!20] (4.5,0.2) rectangle (5.5,-0.2-3*.35);
			\draw [dashed] (4.5,-0.2-1*.35) -- (5.5,-0.2-1*.35);
			\filldraw [fill=yellow!20!blue!20] (6.0,0.2) rectangle (7.0,-0.2-3*.35);
			\draw [dashed] (6.0,-0.2-1*.35) -- (7.0,-0.2-1*.35);
			\filldraw [fill=yellow!20!blue!20] (7.5,0.2) rectangle (8.5,-0.2-3*.35);
			\draw [dashed] (7.5,-0.2-1*.35) -- (8.5,-0.2-1*.35);
			
			\filldraw [fill=yellow!20!purple!20] (9.75,0.22) rectangle (11.25,-0.22-7*.35);
			\filldraw [fill=yellow!20!purple!20] (10.0,0.2) rectangle (11.0,-0.2-3*.35);
			\draw [dashed] (10.0,-0.2-1*.35) -- (11.0,-0.2-1*.35);
			
			\filldraw [fill=yellow!20!purple!20] (12.25,0.2) rectangle (13.75,-0.22-7*.35);
			\filldraw [fill=yellow!20!purple!20] (12.5,-0.2-3*.35) rectangle (13.5,-0.2-7*.35);
			\draw [dashed] (12.5,-0.2-5*.35) -- (13.5,-0.2-5*.35);
			
			
			
			\draw (4.5,-0.2- 7*.35) -- (5.5,-0.2- 7*.35);
			\draw (4.5,-0.2-11*.35) -- (5.5,-0.2-11*.35);
			\draw (4.5,-0.2-15*.35) -- (5.5,-0.2-15*.35);
			\draw (4.5,-0.2-19*.35) -- (5.5,-0.2-19*.35);
			\draw (4.5,-0.2-23*.35) -- (5.5,-0.2-23*.35);
			\draw (4.5,-0.2-27*.35) -- (5.5,-0.2-27*.35);
			\draw (6.0,-0.2- 7*.35) -- (7.0,-0.2- 7*.35);
			\draw (6.0,-0.2-11*.35) -- (7.0,-0.2-11*.35);
			\draw (6.0,-0.2-15*.35) -- (7.0,-0.2-15*.35);
			\draw (6.0,-0.2-19*.35) -- (7.0,-0.2-19*.35);
			\draw (6.0,-0.2-23*.35) -- (7.0,-0.2-23*.35);
			\draw (6.0,-0.2-27*.35) -- (7.0,-0.2-27*.35);
			\draw (7.5,-0.2- 7*.35) -- (8.5,-0.2- 7*.35);
			\draw (7.5,-0.2-11*.35) -- (8.5,-0.2-11*.35);
			\draw (7.5,-0.2-15*.35) -- (8.5,-0.2-15*.35);
			\draw (7.5,-0.2-19*.35) -- (8.5,-0.2-19*.35);
			\draw (7.5,-0.2-23*.35) -- (8.5,-0.2-23*.35);
			\draw (7.5,-0.2-27*.35) -- (8.5,-0.2-27*.35);
			
			\draw (9.75,-0.2-15*.35) -- (11.25,-0.2-15*.35);
			\draw (9.75,-0.2-23*.35) -- (11.25,-0.2-23*.35);
			
			\draw (12.25,-0.2-15*.35) -- (13.75,-0.2-15*.35);
			\draw (12.25,-0.2-23*.35) -- (13.75,-0.2-23*.35);
			
			\foreach \y in {0,1,...,15}
				\node at (0,\y*-0.35) {\texttt0};
			\foreach \y in {16,17,...,31}
				\node at (0,\y*-0.35) {\texttt1};
			\foreach \y in {0,1,...,7,16,17,...,23}
				\node at (0.5,\y*-0.35) {\texttt0};
			\foreach \y in {8,9,...,15,24,25,...,31}
				\node at (0.5,\y*-0.35) {\texttt1};
			\foreach \y in {0,1,...,3,8,9,...,11,16,17,...,19,24,25,...,27}
				\node at (1.0,\y*-0.35) {\texttt0};
			\foreach \y in {4,5,...,7,12,13,...,15,20,21,...,23,28,29,...,31}
				\node at (1.0,\y*-0.35) {\texttt1};
			\foreach \y in {0,1,4,5,8,9,12,13,16,17,20,21,24,25,28,29}
				\node at (1.5,\y*-0.35) {\texttt0};
			\foreach \y in {2,3,6,7,10,11,14,15,18,19,22,23,26,27,30,31}
				\node at (1.5,\y*-0.35) {\texttt1};
			\foreach \y in {0,2,...,30}
				\node at (2.0,\y*-0.35) {\texttt0};	
			\foreach \y in {1,3,...,31}
				\node at (2.0,\y*-0.35) {\texttt1};

			\foreach \y in {0,2,...,30}
				\node at (3.5,\y*-0.35) {\texttt1};
			\foreach \y in {1,3,...,31}
				\node at (3.5,\y*-0.35) {\texttt0};
			
			\foreach \z in {0,4,8,12,16,20,24,28}
			{
				\node at (5.0,\z*-0.35+1*-0.35) {\texttt0};
				\foreach \y in {0,2,3}
					\node at (5.0,\z*-0.35+\y*-0.35) {\texttt1};
			}
				
			\foreach \z in {0,8,16,24}
			{
				\node at (6.5,\z*-0.35) {\texttt0};
				\foreach \y in {1,2,3}
					\node at (6.5,\y*-0.35+\z*-0.35) {\texttt1};
			}
			\foreach \z in {4,12,20,28}
			{
				\node at (6.5,\z*-0.35) {\texttt0};
				\foreach \y in {1,2,3}
					\node at (6.5,\y*-0.35+\z*-0.35) {\texttt1};
			}
			
			\foreach \z in {0,4,8,12,16,20,24,28}
			{
				\node at (8.0,\z*-0.35+2*-0.35) {\texttt0};
				\foreach \y in {0,1,3}
					\node at (8.0,\z*-0.35+\y*-0.35) {\texttt1};
			}
			
			\foreach \z in {0,8,16,24}
			{
				\node at (10.5,\z*-0.35+2*-0.35) {\texttt0};
				\foreach \y in {0,1,3,4,5,6,7}
					\node at (10.5,\z*-0.35+\y*-0.35) {\texttt1};
			}

				
			\foreach \z in {0,8,16,24}
			{
				\node at (13.0,\z*-0.35+6*-0.35) {\texttt0};
				\foreach \y in {0,1,2,3,4,5,7}
					\node at (13.0,\z*-0.35+\y*-0.35) {\texttt1};
			}

		\end{tikzpicture}
	\end{center}
	\caption{\bf{Patterns of Clauses With Negation}\label{NegatedClausePatterns}}
\end{PageFigure}

\begin{PageFigure}
	\begin{center}
		\begin{tikzpicture}
			\filldraw [fill=blue!20, draw=blue] (-3.5,1.5) rectangle (11.5,-4.5);
			\node at (4,1) {\Large{For Clauses of the Form $x_{a_0} \lor x_{a_1} \lor \dots \lor x_{a_z}$}};
			\draw (-1,0.75) -- (9, 0.75);
			\node at (0, 0  ) [right] {$g(\overline{x_{a_i}})=
				\left(\prod_{k=0}^{a_i-1}(1+x^{2^k})\right)$};
			\node at (0,-0.5) [right] {\dots};
			\node at (0,-1.0) [right] {\textsc{If $(h = \overline{a_i})$ for some $\overline{a_i}$}};
			\node at (1,-1.5) [right] {Result = Result$\cdot\left(x^{2^{a_i}}\right) +
				g(\overline{x_{a_i}})$};
			\node at (0,-2.0) [right] {\textsc{Else If $(h = a_i)$ for some $a_i$}};
			\node at (1,-2.5) [right] {Result = Result + $g(x_h)$};
			\node at (0,-3.0) [right] {\textsc{Else}};
			\node at (1,-3.5) [right] {Result = Result$ \cdot (1 + x^{2^h})$};
			\node at (0,-4.0) [right] {\dots};
					
		\end{tikzpicture}
	\end{center}
	\caption{\bf{Negated Clause Polynomial Construction Algorithm}\label{NegatedClauseAlgorithm}}
\end{PageFigure}


\begin{PageFigure}
	\begin{center}
		\begin{tikzpicture}
			\filldraw[fill=blue!20,draw=blue] (-2.5,1.5) rectangle (12.5,-2);
			\node at (5.0, 1  ) {\large{$(x_0 \lor \overline{x_1})$ From $x_0, x_1, x_2$}};
			\draw (2.8,.70) -- (7.2,.70);
			\node at (0  , 0  ) [right] {$g(x_0) = x^{2^0} = 1x^1$};
			\node at (4.5, 0  ) [right] {$g(\overline{x_1}) =
				\left(1 + x^{2^0}\right) = 1x^0 + 1x^1$};
			\node at (0  ,-0.75) [right] {$f(x_0 \lor x_1) =
				\left(\left((1x^1)x^{2^1}\right)+(1x^0+1x^1)\right)\cdot
				\left(1+x^{2^2}\right)$};
			\node at (0  ,-1.50) [right] {$f(x_0 \lor x_1) = {\mbox{\Large1}}x^1 + {\mbox{\Large1}}x^1 +
				{\mbox{\Large0}}x^2 + {\mbox{\Large1}}x^3 + {\mbox{\Large1}}x^4 + {\mbox{\Large1}}x^5 +
				{\mbox{\Large0}}x^6 + {\mbox{\Large1}}x^7$};
			
			\filldraw[fill=purple!20,draw=purple] (-2.5,-2.25) rectangle (12.5,-6.25);
			\node at (4.95,-2.75) {\large{$(x_0 \lor x_1 \lor x_2)$ From $x_0, x_1, x_2$}};
			\draw (2.3,-3.05) -- (7.7,-3.05);
			\node at (-1.5,-3.75) [right] {$g(x_0) = x^{2^0} = 1x^1$};
			\node at ( 3.0,-3.75) [right] {$g(\overline{x_1}) =
				\left(1 + x^{2^0}\right) = 1x^0 + 1x^1$};
			\node at (-1.5,-4.25) [right] {$g(x_2) = \left(1 + x^{2^0}\right)\left(1 + x^{2^1}\right)
				x^{2^2} = 1x^4 + 1x^5 + 1x^6 + 1x^7$};
			\node at (-1.5,-5.00) [right] {$f(x_0 \lor x_1 \lor x_2) =
				\left(\left((1x^1)x^{2^1}\right)+(1x^0+1x^1)\right)\cdot\left(1+x^{2^2}\right) +
				(1x^4 + 1x^5 + 1x^6 + 1x^7)$};
			\node at (-1.5,-5.75) [right] {$f(x_0 \lor x_1 \lor x_2) = {\mbox{\Large1}}x^0 +
				{\mbox{\Large1}}x^1 + {\mbox{\Large0}}x^2 + {\mbox{\Large1}}x^3 + {\mbox{\Large1}}x^4 +
				{\mbox{\Large1}}x^5 + {\mbox{\Large1}}x^6 + {\mbox{\Large1}}x^7$};
		\end{tikzpicture}
	\end{center}
	\caption{\bf{Clause Polynomial Examples With Negation}\label{NegatedClauseExamples}}
\end{PageFigure}

\pagebreak

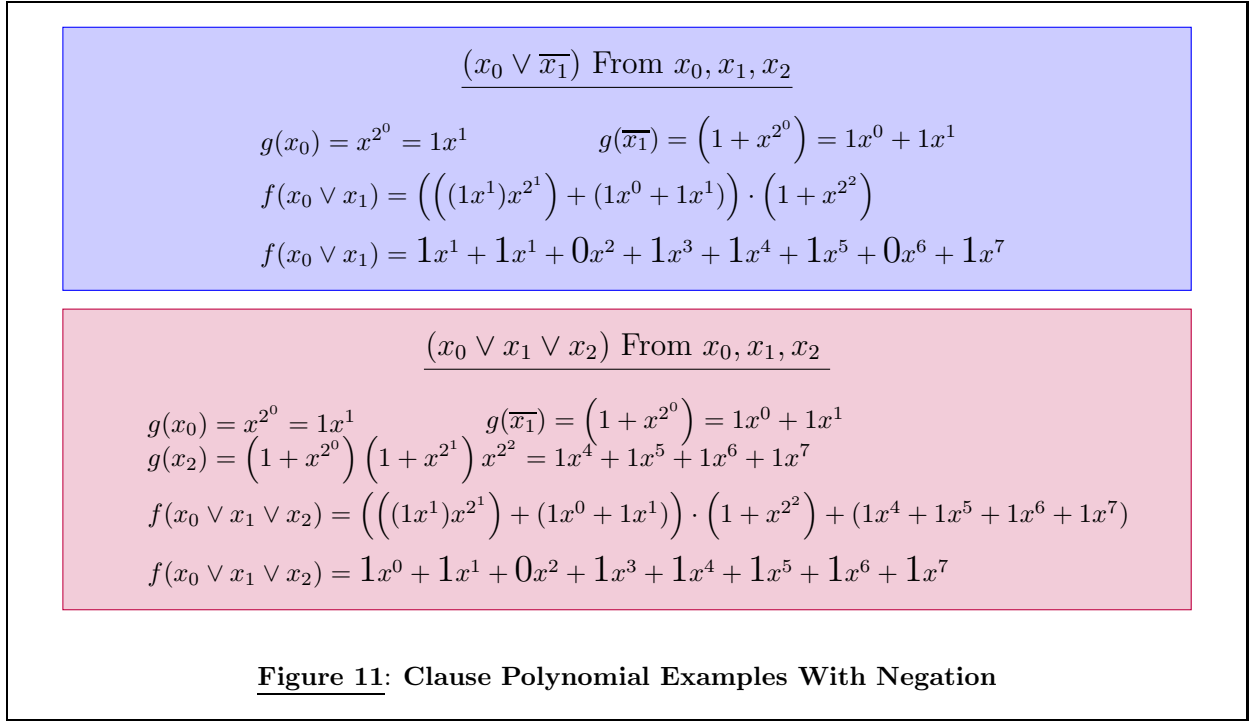
Figure \ref{NegatedClausePatterns}, on the previous page, displays clause patterns with negated variables.  As can be noted by the negated variable $x_0$ near the left, it is just a reversed version of ones and zeros, compared to the non-negated $x_0$ displayed just to the left of it.  In other words, just as negation reverses true and false values in Boolean algebra, negation reverses zeros and ones in clause polynomials.  In fact, this is how all single-variable clauses are negated;  simply swap the zeros and ones.  There is actually a very simple and effective way to do this - the negated clauses are the same as the regular clauses except for the fact that they are not multiplied by a power of $x^{2^m}$, as the regular clauses are.  This is the equivalent as removing this term from Equation \ref{XmEquation}, hence the resulting equation:

\defeq{NegatedXmEquation}
f(\overline{x_m})=\left(\prod_{k=0}^{m-1}(1+x^{2^k})\right)\left(\prod_{k=m+1}^{n}(1+x^{2^k})\right)
\eeq

Again, this is for a system of $n+1$ variables, and $n$ can be adjusted accordingly for the size of the problem.

Next, attention can be turned to the new patterns observed for multiple-variable clauses.  Figure \ref{NegatedClausePatterns} shows the negated values of $\overline{x_0}$ next to the combined clause $\overline{x_0} \lor x_1$.  Note that the pattern of ones coming from $x_1$ has not changed in any way; however, the repeating pattern of zeros and ones coming from $\overline{x_0}$ has been switched, compared to the non-negated equation $x_0 \lor x_1$ beside it on the right.  So really, it is becoming evident that negating variables only swaps the pattern of ones that the variable creates.

This is more evident if the clause $x_0 \lor \overline{x_1}$ is observed beside it.  Here the pattern of ones coming from the variable $x_1$ has changed, since $x_1$ has been negated.  So again, there is evidence that the pattern of ones coming from a variable is simply ``exchanged'' or placed elsewhere.  It can be seen, though, that this does not change the pattern for $x_2$, as evidenced in the equation $x_0 \lor \overline{x_1} \lor x_2$ beside it.

Finally, it's seen that swapping the pattern of ones may entail moving a smaller group of ones and zeros.  This is evidenced in the final clause in the figure.  Here, the variable $x_2$ has been negated, which causes a shift in the largest pattern of ones, which is due to $x_2$.
  
Two examples have been worked out in slight detail in Figure \ref{NegatedClauseExamples}.  Here the pattern is mostly the same, with the exception that negating variables causes some values to be swapped.  This is the equivalent of moving the pattern of ones.  The adjustments to the algorithm are shown in Figure \ref{NegatedClauseAlgorithm} on the previous page.

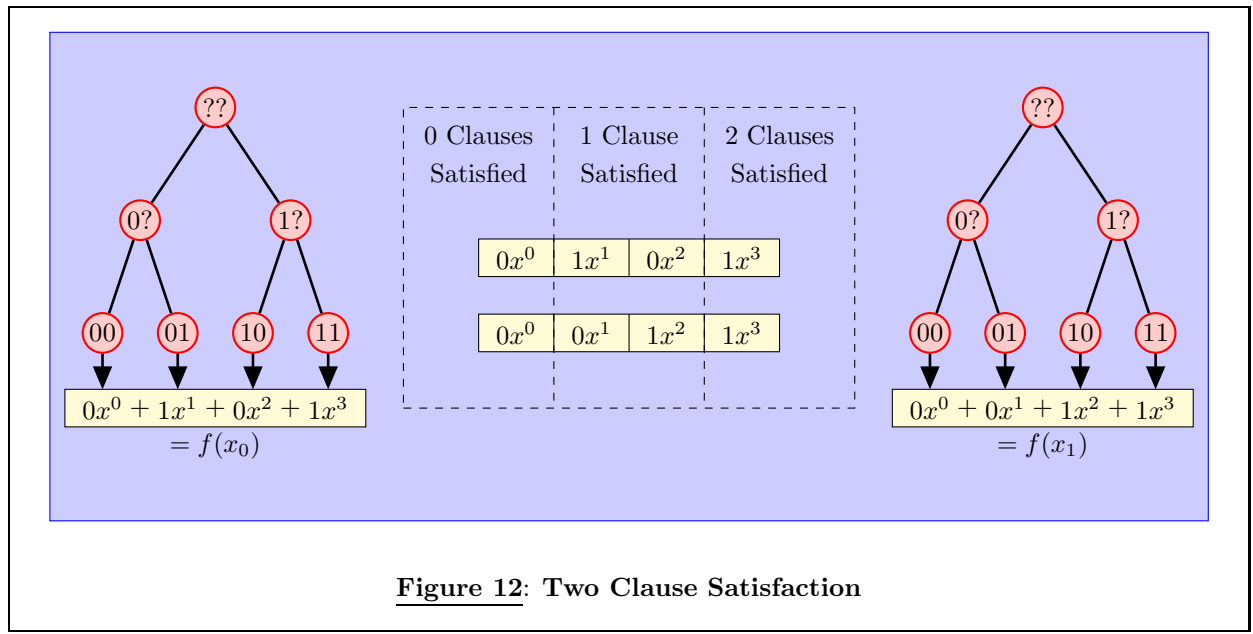
\begin{PageFigure}
	\begin{center}
		\begin{tikzpicture}
			\tikzstyle{N2}=[circle,draw=red,fill=red!20,thick, inner sep=1pt,minimum size=15];
			
			\filldraw [fill=blue!20, draw=blue] (-2.2,1) rectangle (13.2,-5.5);
			
			\node (1xx) at ( 0  , 0  ) [N2] {??};
			\node (10x) at (-1  ,-1.5) [N2] {0?};
			\node (11x) at ( 1  ,-1.5) [N2] {1?};
			\node (100) at (-1.5,-3  ) [N2] {00}; 
			\node (101) at (-0.5,-3  ) [N2] {01};
			\node (110) at ( 0.5,-3  ) [N2] {10};
			\node (111) at ( 1.5,-3  ) [N2] {11};
			
			\draw [line width = 1] (1xx) -- (10x);
			\draw [line width = 1] (1xx) -- (11x);
			\draw [line width = 1] (10x) -- (100);
			\draw [line width = 1] (10x) -- (101);
			\draw [line width = 1] (11x) -- (110);
			\draw [line width = 1] (11x) -- (111);
			
			\filldraw [fill=yellow!20] (-2,-3.75) rectangle (2,-4.25);
			
			\draw [line width = 1,-triangle 45] (100) -- (-1.5, -3.75);
			\draw [line width = 1,-triangle 45] (101) -- (-0.5, -3.75);
			\draw [line width = 1,-triangle 45] (110) -- ( 0.5, -3.75);
			\draw [line width = 1,-triangle 45] (111) -- ( 1.5, -3.75);
			
			\node at (-1.5,-4) {$0x^0$};
			\node at (-0.5,-4) {$1x^1$};
			\node at ( 0.5,-4) {$0x^2$};
			\node at ( 1.5,-4) {$1x^3$};
			
			\foreach \x in {-1,0,1}
				\node at (\x,-4) {$+$};
			\node at (0,-4.5) {$= f(x_0)$};
			
			\node (2xx) at (11  , 0  ) [N2] {??};
			\node (20x) at (10  ,-1.5) [N2] {0?};
			\node (21x) at (12  ,-1.5) [N2] {1?};
			\node (200) at ( 9.5,-3  ) [N2] {00}; 
			\node (201) at (10.5,-3  ) [N2] {01};
			\node (210) at (11.5,-3  ) [N2] {10};
			\node (211) at (12.5,-3  ) [N2] {11};
			\draw [line width = 1] (2xx) -- (20x);
			\draw [line width = 1] (2xx) -- (21x);
			\draw [line width = 1] (20x) -- (200);
			\draw [line width = 1] (20x) -- (201);
			\draw [line width = 1] (21x) -- (210);
			\draw [line width = 1] (21x) -- (211);
			
			\filldraw [fill=yellow!20] ( 9,-3.75) rectangle (13,-4.25);
			
			\draw [line width = 1,-triangle 45] (200) -- ( 9.5, -3.75);
			\draw [line width = 1,-triangle 45] (201) -- (10.5, -3.75);
			\draw [line width = 1,-triangle 45] (210) -- (11.5, -3.75);
			\draw [line width = 1,-triangle 45] (211) -- (12.5, -3.75);
			
			\node at ( 9.5,-4) {$0x^0$};
			\node at (10.5,-4) {$0x^1$};
			\node at (11.5,-4) {$1x^2$};
			\node at (12.5,-4) {$1x^3$};
			
			\foreach \x in {10,11,12}
				\node at (\x,-4) {$+$};
			\node at (11,-4.5) {$= f(x_1)$};
			
			\filldraw [fill=yellow!20] (3.5,-1.75) rectangle (7.5,-2.25);
			\foreach \x in {4.5,5.5,6.5}
			{
				\draw (\x,-1.75) -- (\x,-2.25);
			}
				
			\node at (4,-2) {$0x^0$};
			\node at (5,-2) {$1x^1$};
			\node at (6,-2) {$0x^2$};
			\node at (7,-2) {$1x^3$};
			
			\filldraw [fill=yellow!20] (3.5,-2.75) rectangle (7.5,-3.25);
			\foreach \x in {4.5,5.5,6.5}
			{
				\draw (\x,-2.75) -- (\x,-3.25);
			}
				
			\node at (4,-3) {$0x^0$};
			\node at (5,-3) {$0x^1$};
			\node at (6,-3) {$1x^2$};
			\node at (7,-3) {$1x^3$};
			
			\draw [dashed] (2.5,0) rectangle (8.5,-4);
			\draw [dashed] (4.5,0) -- (4.5,-4);
			\draw [dashed] (6.5,0) -- (6.5,-4);
			
			\node at (3.5,-0.37) {0 Clauses};
			\node at (5.5,-0.37) {1 Clause};
			\node at (7.5,-0.37) {2 Clauses};
			\node at (3.5,-0.87) {Satisfied};
			\node at (5.5,-0.87) {Satisfied};
			\node at (7.5,-0.87) {Satisfied};
			
		\end{tikzpicture}
	\end{center}
	\caption{\bf{Two Clause Satisfaction}\label{TwoClauseSatisfaction}}
\end{PageFigure}
\pagebreak

\Section{Two Clause Problems}
Now that sufficient background has been presented, the major ideas behind the algorithm can be introduced.  Two clause problems are some of the simplest cases, yet they allow the fundamental ideas to be introduced and studied.

Figure \ref{TwoClauseSatisfaction} shows the four basic possible combinations between two clauses.  If a clause polynomial is considered with any amount of terms (powers of $x$), there are only two possible values (known in mathematics as coefficients) that can be associated with each term; zero or one.  If two clauses are considered, there are two possible values for the coefficient of the first polynomial, and two possible values for the coefficient of the second polynomial, for a total of four different combinations.

The figure shows two clause polynomials and the trees associated with them.  In the middle, the four possible combinations can be seen.  Remembering that the coefficients represent satisfaction for a particular truth assignment, it's possible to note the total satisfaction for two clauses considered simultaneously.  When both clauses have a zero coefficient, that indicates that neither clause will satisfy the original question of satisfaction for that particular truth assignment.  Looking through the cluase trees, it's seen that this assignment corresponds to $(x_0=\mbox{false}, x_1=\mbox{false})$.  So this truth assignment won't satisfy any of the clauses.  On the other hand, the two truth assignments in the middle of the figure each satisfy one clause.  The truth assignment $(x_0=\mbox{true}, x_1=\mbox{true})$ is seen to satisfy both clauses, and so it is a solution to the original problem.  Again, the reason it satisfies both clauses is that it has a one for both coefficients, thus signifying that both clauses are satisfied.

One important thing to note here is that between two clauses, there are really only three possible satisfaction results.  Either zero, one, or both of the clauses are satisfied (for any particular truth assignment).  One general idea that, although perhaps trivial, will be important is that there are really only three types of satisfaction between two clauses.  This basic concept will be extended to see that there are really only $n+1$ possible types of satisfaction between $n$ clauses.

The fundamental idea here is that the problems can be simplified so that large problems can really be dealt with by simply considering the types of satisfaction, which are fairly simple.  For two clause problems, it will really only be necessary to deal with the three types of satisfaction, and interactions between the different types of satisfaction.

It will be shown how operations of multiplication and addition can be used to work with the three different types of satisfaction, and to solve questions about them.  This will be essentially simpler and in some ways necessary to overcome the complications of working with all of the possibilities between two or more clauses.

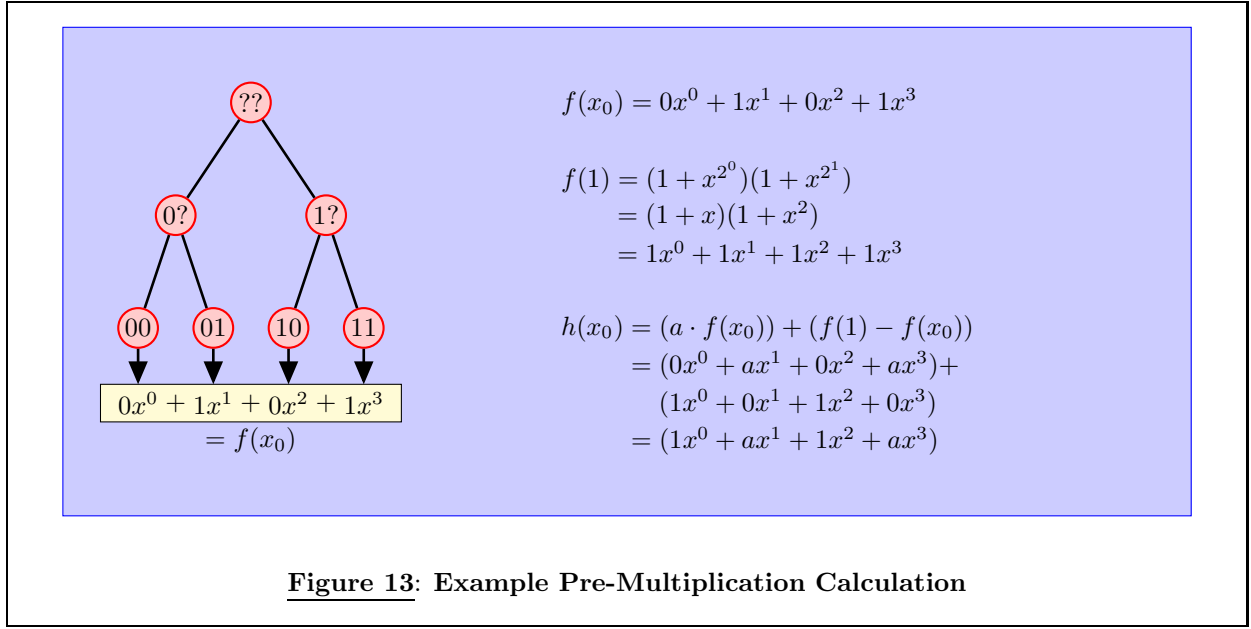
\begin{PageFigure}
	\begin{center}
		\begin{tikzpicture}
			\tikzstyle{N2}=[circle,draw=red,fill=red!20,thick, inner sep=1pt,minimum size=15];
			
			\filldraw [fill=blue!20, draw=blue] (-2.5,1) rectangle (12.5,-5.5);
			
			\node (1xx) at ( 0  , 0  ) [N2] {??};
			\node (10x) at (-1  ,-1.5) [N2] {0?};
			\node (11x) at ( 1  ,-1.5) [N2] {1?};
			\node (100) at (-1.5,-3  ) [N2] {00}; 
			\node (101) at (-0.5,-3  ) [N2] {01};
			\node (110) at ( 0.5,-3  ) [N2] {10};
			\node (111) at ( 1.5,-3  ) [N2] {11};
			
			\draw [line width = 1] (1xx) -- (10x);
			\draw [line width = 1] (1xx) -- (11x);
			\draw [line width = 1] (10x) -- (100);
			\draw [line width = 1] (10x) -- (101);
			\draw [line width = 1] (11x) -- (110);
			\draw [line width = 1] (11x) -- (111);
			
			\filldraw [fill=yellow!20] (-2,-3.75) rectangle (2,-4.25);
			
			\draw [line width = 1,-triangle 45] (100) -- (-1.5, -3.75);
			\draw [line width = 1,-triangle 45] (101) -- (-0.5, -3.75);
			\draw [line width = 1,-triangle 45] (110) -- ( 0.5, -3.75);
			\draw [line width = 1,-triangle 45] (111) -- ( 1.5, -3.75);
			
			\node at (-1.5,-4) {$0x^0$};
			\node at (-0.5,-4) {$1x^1$};
			\node at ( 0.5,-4) {$0x^2$};
			\node at ( 1.5,-4) {$1x^3$};
			
			\foreach \x in {-1,0,1}
				\node at (\x,-4) {$+$};
			\node at (0,-4.5) {$= f(x_0)$};
			
			\node at (4.0 , 0  ) [right] {$f(x_0) = 0x^0 + 1x^1 + 0x^2 + 1x^3$};
			\node at (4.0 ,-1  ) [right] {$f(1) = (1 + x^{2^0})(1 + x^{2^1})$};
			\node at (4.75,-1.5) [right] {$= (1+x)(1+x^2)$};
			\node at (4.75,-2  ) [right] {$= 1x^0 + 1x^1 + 1x^2 + 1x^3$};
			\node at (4.0 ,-3  ) [right] {$h(x_0) = \left(a \cdot f(x_0)\right) +
				\left(f(1) - f(x_0)\right)$};
			\node at (4.93,-3.5) [right] {$=(0x^0 + ax^1 + 0x^2 + ax^3) +$};
			\node at (5.30,-4  ) [right] {$(1x^0 + 0x^1 + 1x^2 + 0x^3)$};
			\node at (4.93,-4.5) [right] {$=(1x^0 + ax^1 + 1x^2 + ax^3)$};

		\end{tikzpicture}
	\end{center}
	\caption{\bf{Example Pre-Multiplication Calculation}\label{MultiplicationPreCalculation}}
\end{PageFigure}

\Subsection{Manipulating Clause\\ Polynomials}
In order to simplify clause calculations into the the equivalence classes of satisfaction, it is necessary to modify the original clause polynomials.

The first modification used is a simple one.  The algorithm will need to change all of the coefficients that are equal to one into coefficients that are equal to $a$.  This is easy; it is just multiplication by $a$.  Here's an example:

\begin{align}
\nonumber f(x) &= 0x^0 + 1x^1 + 0x^2 + 1x^3\\
\nonumber a \cdot f(x) &= 0x^0 + a  x^1 + 0x^2 + a x^3
\end{align}

The next modification is a bit tougher; it is exchanging the one and zero coefficients.  This can be done by subtracting the original function from a function of ones.  This inverts the bits, since the ones are subtracted from ones to become zeros, and the the zeros are subtracted from ones to become ones.  However, the function of all ones is needed.  This function is, for a system of $V$ variables:

\defeq{OnesEquation}
f(1) = \prod_{k=0}^{V-1}{(1+x^{2^k})}
\eeq

Note that:

\defeq{OnesEquality}
\nonumber f(1) = 1x^0 + 1x^2 + 1x^3 + \dots + 1x^{2^V-1}
\eeq

This completes the requirements for multiplication.  The algorithm uses this knowledge to change the one coefficients into $a$ coefficients and the zero coefficients into one coefficients.  Then multiplication can be performed, as will be seen.

Figure \ref{MultiplicationPreCalculation} exhibits the steps taken before a function is ready for multiplication.  $f(x_0)$ is given through a tree, before multiplication.  The ones must be transformed into $a$s, and the zeros must be transformed into ones.  Note the final result, $h(x_0)$.  It is the finished calculation, ready for multiplication.

To do this, the function of ones for a two-variable system must be calculated.  Then the original function can be multiplied by $a$ to transform the one coefficients, and it is also subtracted from the function of ones (seperately) to transform the zero coefficients.  These are added together for the final result.

Note that this procedure works for any clause in any system with any amount of variables.  It simply prepares the clause polynomials for special processing, or interactions, with other clause polynomials.  This may not seem like much, but it greatly simplifies the system.

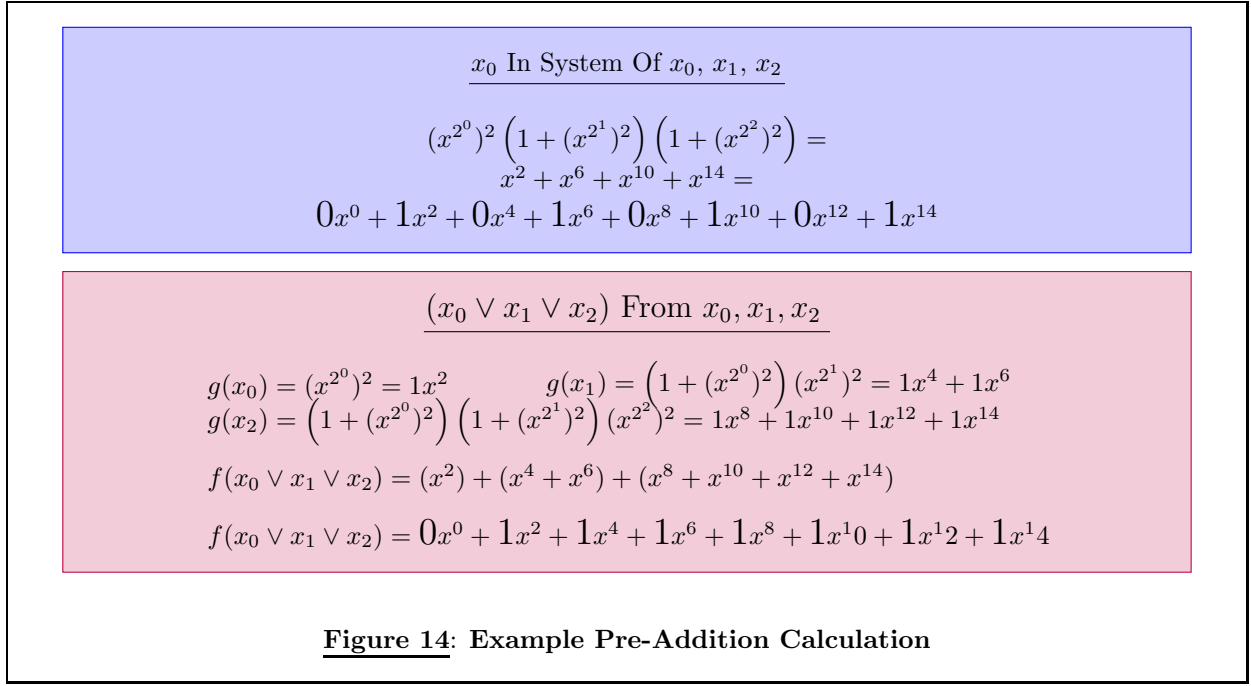
\begin{PageFigure}
	\begin{center}
		\begin{tikzpicture}
			\filldraw [fill=blue!20,draw=blue] (-7.5,1) rectangle (7.5,-2);
			\node at (0, 0.5) {$x_0$ In System Of $x_0$, $x_1$, $x_2$};
			\draw (-2.1, 0.25) -- (2.1, 0.25);
			\node at (0,-0.5) {$(x^{2^0})^2\left(1 + (x^{2^1})^2\right)\left(1 + (x^{2^2})^2\right) =$};
			\node at (0,-1  ) {$x^2 + x^6 + x^{10} + x^{14} =$};
			\node at (0,-1.5) {$\mbox{\Large0}x^0 + \mbox{\Large1}x^2 + \mbox{\Large0}x^4 +
				\mbox{\Large1}x^6 + \mbox{\Large0}x^8 + \mbox{\Large1}x^{10} +
				\mbox{\Large0}x^{12} + \mbox{\Large1}x^{14}$};
				
			\filldraw[fill=purple!20,draw=purple] (-2.5-5,-2.25) rectangle (12.5-5,-6.25);
			\node at (4.95-5,-2.75) {\large{$(x_0 \lor x_1 \lor x_2)$ From $x_0, x_1, x_2$}};
			\draw (2.3-5,-3.05) -- (7.7-5,-3.05);
			\node at (0  -5.7,-3.75) [right] {$g(x_0) = (x^{2^0})^2 = 1x^2$};
			\node at (4.5-5.7,-3.75) [right] {$g(x_1) = \left(1 + (x^{2^0})^2\right)
				(x^{2^1})^2 = 1x^4 + 1x^6$};
			\node at (0  -5.7,-4.25) [right] {$g(x_2) = \left(1 + (x^{2^0})^2\right)
				\left(1 + (x^{2^1})^2\right)(x^{2^2})^2 = 1x^8 + 1x^{10} + 1x^{12} + 1x^{14}$};
			\node at (0  -5.7,-5.00) [right] {$f(x_0 \lor x_1 \lor x_2) = (x^2) + (x^4 + x^6) +
				(x^8 + x^{10} + x^{12} + x^{14})$};
			\node at (0  -5.7,-5.75) [right] {$f(x_0 \lor x_1 \lor x_2) = {\mbox{\Large0}}x^0 +
				{\mbox{\Large1}}x^2 + {\mbox{\Large1}}x^4 + {\mbox{\Large1}}x^6 + {\mbox{\Large1}}x^8 +
				{\mbox{\Large1}}x^10 + {\mbox{\Large1}}x^12 + {\mbox{\Large1}}x^14$};
				
		\end{tikzpicture}
	\end{center}
	\caption{\bf{Example Pre-Addition Calculation}\label{AdditionPreCalculation}}
\end{PageFigure}

\pagebreak
Addition requires modifications too.  However, these modifications are of a different variety than those of multiplication.  Essentially, the power of $x$ needs to be doubled.

Doubling the power of $x$ can actually be fairly simple.  Figure \ref{AdditionPreCalculation} displays a retake of the creation of clause polynomials.  The first part of the example is really just Figure \ref{VariableConstruction} on page \pageref{VariableConstruction}, modified for doubling the power of $x$.  Note that each power of $x$, once it is essentially calculated, is doubled.  The result is that all of the powers of $x$ are multiplied by two in the finished clause polynomial.

The second part of the figure is also a redo, this time of a multiple variable clause taken from Figure \ref{ClauseExamples} on page \pageref{ClauseExamples}.  Again, the main idea is that the powers of the variable $x$ is doubled.

The actual operation of addition is also performed on one special value, which comes in part from the ideas of multiplication.  A constant is added to each coefficient of the polynomials.  This constant is the same value for all coefficients, so the function of ones once again becomes useful.  Unfortunately, in its originally derived form,  the powers of $x$ are not the same as the powers of $x$ used in addition.  So once again, the same principles are used to double the powers of $x$ for the function of ones.  It is really as simple to do this as altering every power of $x$ in the original equation (Equation \ref{OnesFunction} on page \pageref{OnesFunction}).  The new formula for the modified function of ones is as follows:

\defeq{AdditionOnes}
f(1 \mbox{\textit{modified}}) = \prod_{k=0}^{V-1}{\left(1 + (x^{2^k})^2\right)}
\eeq

These operations will be very useful in the material ahead.

\begin{PageFigure}
	\begin{center}
		\begin{tikzpicture}
			\tikzstyle{N2}=[circle,draw=red,fill=red!20,thick, inner sep=1pt,minimum size=15];
			
			\filldraw [fill=blue!20, draw=blue] (-2.2,1) rectangle (13.2,-5.5);
			
			\node (1xx) at ( 0  , 0  ) [N2] {??};
			\node (10x) at (-1  ,-1.5) [N2] {0?};
			\node (11x) at ( 1  ,-1.5) [N2] {1?};
			\node (100) at (-1.5,-3  ) [N2] {00}; 
			\node (101) at (-0.5,-3  ) [N2] {01};
			\node (110) at ( 0.5,-3  ) [N2] {10};
			\node (111) at ( 1.5,-3  ) [N2] {11};
			
			\draw [line width = 1] (1xx) -- (10x);
			\draw [line width = 1] (1xx) -- (11x);
			\draw [line width = 1] (10x) -- (100);
			\draw [line width = 1] (10x) -- (101);
			\draw [line width = 1] (11x) -- (110);
			\draw [line width = 1] (11x) -- (111);
			
			\filldraw [fill=yellow!20] (-2,-3.75) rectangle (2,-4.25);
			
			\draw [line width = 1,-triangle 45] (100) -- (-1.5, -3.75);
			\draw [line width = 1,-triangle 45] (101) -- (-0.5, -3.75);
			\draw [line width = 1,-triangle 45] (110) -- ( 0.5, -3.75);
			\draw [line width = 1,-triangle 45] (111) -- ( 1.5, -3.75);
			
			\node at (-1.5,-4) {$0x^0$};
			\node at (-0.5,-4) {$1x^1$};
			\node at ( 0.5,-4) {$0x^2$};
			\node at ( 1.5,-4) {$1x^3$};
			
			\foreach \x in {-1,0,1}
				\node at (\x,-4) {$+$};
			\node at (0,-4.5) {$= f(x_0)$};
			
			\node (2xx) at (11  , 0  ) [N2] {??};
			\node (20x) at (10  ,-1.5) [N2] {0?};
			\node (21x) at (12  ,-1.5) [N2] {1?};
			\node (200) at ( 9.5,-3  ) [N2] {00}; 
			\node (201) at (10.5,-3  ) [N2] {01};
			\node (210) at (11.5,-3  ) [N2] {10};
			\node (211) at (12.5,-3  ) [N2] {11};
			\draw [line width = 1] (2xx) -- (20x);
			\draw [line width = 1] (2xx) -- (21x);
			\draw [line width = 1] (20x) -- (200);
			\draw [line width = 1] (20x) -- (201);
			\draw [line width = 1] (21x) -- (210);
			\draw [line width = 1] (21x) -- (211);
			
			\filldraw [fill=yellow!20] ( 9,-3.75) rectangle (13,-4.25);
			
			\draw [line width = 1,-triangle 45] (200) -- ( 9.5, -3.75);
			\draw [line width = 1,-triangle 45] (201) -- (10.5, -3.75);
			\draw [line width = 1,-triangle 45] (210) -- (11.5, -3.75);
			\draw [line width = 1,-triangle 45] (211) -- (12.5, -3.75);
			
			\node at ( 9.5,-4) {$0x^0$};
			\node at (10.5,-4) {$0x^1$};
			\node at (11.5,-4) {$1x^2$};
			\node at (12.5,-4) {$1x^3$};
			
			\foreach \x in {10,11,12}
				\node at (\x,-4) {$+$};
			\node at (11,-4.5) {$= f(x_1)$};
			
			\fill [fill=purple!20] (4,-1  ) rectangle (8,-1.5);
			\fill [fill=purple!20] (3,-1.5) rectangle (4,-3.5);
			\fill [fill=red!20]    (3,-1  ) rectangle (4,-1.5);
			\fill [fill=gray!50]   (4,-1.5) rectangle (8,-3.5);
			\foreach \x in {4,5,6,7}
				\fill [fill=yellow!20] (\x,\x*-0.5+.5) rectangle (\x+1,\x*-0.5);
			\draw [xstep=1,ystep=.5](3,-3.5) grid (8,-1);
			
			\node at (3.5,-1.25) {$\cdot$};
			\node at (4.5,-1.25) {$0x^0$};
			\node at (5.5,-1.25) {$1x^1$};
			\node at (6.5,-1.25) {$0x^2$};
			\node at (7.5,-1.25) {$1x^3$};
			
			\node at (3.5,-1.75) {$0x^0$};
			\node at (3.5,-2.25) {$0x^0$};
			\node at (3.5,-2.75) {$1x^0$};
			\node at (3.5,-3.25) {$1x^0$};
			
			\node at (4.5,-1.75) {$0x^0$};
			\node at (5.5,-1.75) {$0x^1$};
			\node at (6.5,-1.75) {$0x^2$};
			\node at (7.5,-1.75) {$0x^3$};
			\node at (4.5,-2.25) {$0x^1$};
			\node at (5.5,-2.25) {$0x^2$};
			\node at (6.5,-2.25) {$0x^3$};
			\node at (7.5,-2.25) {$0x^4$};
			\node at (4.5,-2.75) {$0x^2$};
			\node at (5.5,-2.75) {$1x^3$};
			\node at (6.5,-2.75) {$0x^4$};
			\node at (7.5,-2.75) {$1x^5$};
			\node at (4.5,-3.25) {$0x^3$};
			\node at (5.5,-3.25) {$1x^4$};
			\node at (6.5,-3.25) {$0x^5$};
			\node at (7.5,-3.25) {$1x^6$};
		\end{tikzpicture}
	\end{center}
	\caption{\bf{Clause Multiplication}\label{ClauseMultiplication}}
\end{PageFigure}

\pagebreak

\Subsection{Multiplication}
It could be said that the algorithm is focused around (arithmetic) multiplication of clause polynomials.

Figure \ref{ClauseMultiplication} details an example.  It begins with two clause polynomials, $f(x_0)$ and $f(x_1)$.  The associated trees are shown that go with the clause polynomials.  In the middle, the clause polynomials are broken into pieces and multiplied piecewise.  Every part of each polynomial is multiplied with every part of the other polynomial.  This corresponds with plain old arithmetic multiplication of two polynomials.  However, an interesting event happens here.  The result that lies along the diagonal (shown in yellow) corresponds with the result $f(x_0 \land x_1)$.  That is, the diagonal represents the corresponding clause polynomial for the result of conjunction.  The idea here is that multiplication of clause polynomials can be used to evaluate the original Boolean equation.

\begin{ColumnFigure}
	\begin{align*}
		\begin{tabular}{|c||c|c|}
			\hline
			$\land$ & False & True \\ \hline
			\hline
			False & False & False \\ \hline
			True & False & True \\ \hline
		\end{tabular} \
		& \quad \begin{tabular}{|c||c|c|}
			\hline
			$\cdot$ & 0 & 1 \\ \hline
			\hline
			0 & 0 & 0 \\ \hline
			1 & 0 & 1 \\ \hline
		\end{tabular} 
	\end{align*}
	\caption{\bf{Operation Equivalence}\label{Equivalence}}
\end{ColumnFigure}

Figure \ref{Equivalence} helps give a partial explanation of why this occurs.  As can be seen from the truth tables, the logical operation of conjunction ($\land$) and the arithmetic operation of multiplication($\cdot$) are equivalent on bit values.  So it's not totally unpredictable that multiplication of a clause polynomial contains results similar to conjunction.  The diagonal in Figure \ref{ClauseMultiplication} contains like terms multiplied by like terms.  Only the coefficients (or numbers attached to the terms) may differ.  The result (along the diagonal) is known in mathematics as the Hadamard product, and the algorithm is concentrated on seperating this diagonal term from the off-diagonal terms.

The reason why the algorithm is so closely associated with the Hadamard product, or the diagonal terms, is that it is essentially the solution to the original problem.  Since it tells which terms are satisfying, the algorithm can simply count the number of satisfying terms along the diagonal.  If there are any satisfying terms, then the original problem can be satisfied.  Otherwise, it can't be satisfied.

So at this point in the ideaology, the original Boolean problem has been transformed from a question about Boolean arithmetic, to a question concerning how to get information about the diagonal (or Hadamard product).

\begin{PageFigure}
	\begin{center}
		\begin{tikzpicture}
		
			\tikzstyle{Circ}=[circle,draw=red,fill=red!20,thick, inner sep=1pt,minimum size=15]
		
			\filldraw [fill=blue!20, draw=blue] (-.87,2.25) rectangle (14.12,-5.25);

			\foreach \x in {0,1,...,3}
			{
				\filldraw [fill=yellow!20, draw=black] (\x*1.5,0) rectangle (\x*1.5+1.25,-0.5);
				
				\filldraw [fill=yellow!20, draw=black] (\x*1.5,-2) rectangle (\x*1.5+1.25,-2.5);
				
			}
			
			\foreach \x in {5,6,...,8}
			{
				\filldraw [fill=yellow!20, draw=black] (\x*1.5,0) rectangle (\x*1.5+1.25,-0.5);
				
				\node (cdot\x) at (\x*1.5 + .63,-1.25) [Circ] {$\cdot$};
				\draw [line width = 1,-triangle 45] (\x*1.5 + .63,-0.5) -- (cdot\x);
				\draw [line width = 1,-triangle 45] (cdot\x) -- (\x*1.5 + .63,-2);
				\filldraw [fill=yellow!20, draw=black] (\x*1.5,-2) rectangle (\x*1.5+1.25,-2.5);
				
				\node (cdot2\x) at (\x*1.5 + .63,-3.25) [Circ] {=};
				\draw [line width = 1,-triangle 45] (\x*1.5 + .63,-2.5) -- (cdot2\x);
				\draw [line width = 1,-triangle 45] (cdot2\x) -- (\x*1.5 + .63,-4);
				\filldraw [fill=yellow!20, draw=black] (\x*1.5,-4) rectangle (\x*1.5+1.25,-4.5);
			}
			
			\node (1topmult) at (.63 + 0*1.5,-.25) {$0\cdot x$};
			\node (2topmult) at (.63 + 1*1.5,-.25) {$1\cdot x$};
			\node (3topmult) at (.63 + 2*1.5,-.25) {$0\cdot x$};
			\node (4topmult) at (.63 + 3*1.5,-.25) {$1\cdot x$};
			
			\node (1botmult) at (.63 + 0*1.5,-2.25) {$0\cdot x$};
			\node (2botmult) at (.63 + 1*1.5,-2.25) {$0\cdot x$};
			\node (3botmult) at (.63 + 2*1.5,-2.25) {$1\cdot x$};
			\node (4botmult) at (.63 + 3*1.5,-2.25) {$1\cdot x$};

			\node (1topadd) at (.63 + 5*1.5,-.25) {$1\cdot x$};
			\node (2topadd) at (.63 + 6*1.5,-.25) {$a\cdot x$};
			\node (3topadd) at (.63 + 7*1.5,-.25) {$1\cdot x$};
			\node (4topadd) at (.63 + 8*1.5,-.25) {$a\cdot x$};
			
			\node (1botadd) at (.63 + 5*1.5,-2.25) {$1\cdot x$};
			\node (2botadd) at (.63 + 6*1.5,-2.25) {$1\cdot x$};
			\node (3botadd) at (.63 + 7*1.5,-2.25) {$a\cdot x$};
			\node (4botadd) at (.63 + 8*1.5,-2.25) {$a\cdot x$};
			
			\node (1resadd) at (.63 + 5*1.5,-4.25) {$1x^2$};
			\node (2resadd) at (.63 + 6*1.5,-4.25) {$ax^2$};
			\node (3resadd) at (.63 + 7*1.5,-4.25) {$ax^2$};
			\node (4resadd) at (.63 + 8*1.5,-4.25) {$a^2x^2$};
			
			
			
			\draw [dashed] (-0.5,2) rectangle (6.25,-3);
			\draw [dashed] (1.37,2) -- (1.37,-3);
			\draw [dashed] (4.37,2) -- (4.37,-3);
			\node at (0.41,1.5) {0 Clauses};
			\node at (0.41,1.0) {Satisfied};
			\node at (2.87,1.5) {1 Clause};
			\node at (2.87,1.0) {Satisfied};
			\node at (5.31,1.5) {2 Clauses};
			\node at (5.31,1.0) {Satisfied};
			
			\draw [dashed] (7.0,2) rectangle (13.75,-5);
			\draw [dashed] (8.87,2) -- (8.87,-5);
			\draw [dashed] (11.87,2) -- (11.87,-5);
			\node at (7.91,1.5) {0 Clauses};
			\node at (7.91,1.0) {Satisfied};
			\node at (10.37,1.5) {1 Clause};
			\node at (10.37,1.0) {Satisfied};
			\node at (12.81,1.5) {2 Clauses};
			\node at (12.81,1.0) {Satisfied};
			
			\draw [line width = 2,-triangle 45] (5.75,-0.25) -- (7.5,-0.25);
			\draw [line width = 2,-triangle 45] (5.75,-2.25) -- (7.5,-2.25);
			
		\end{tikzpicture}
	\end{center}
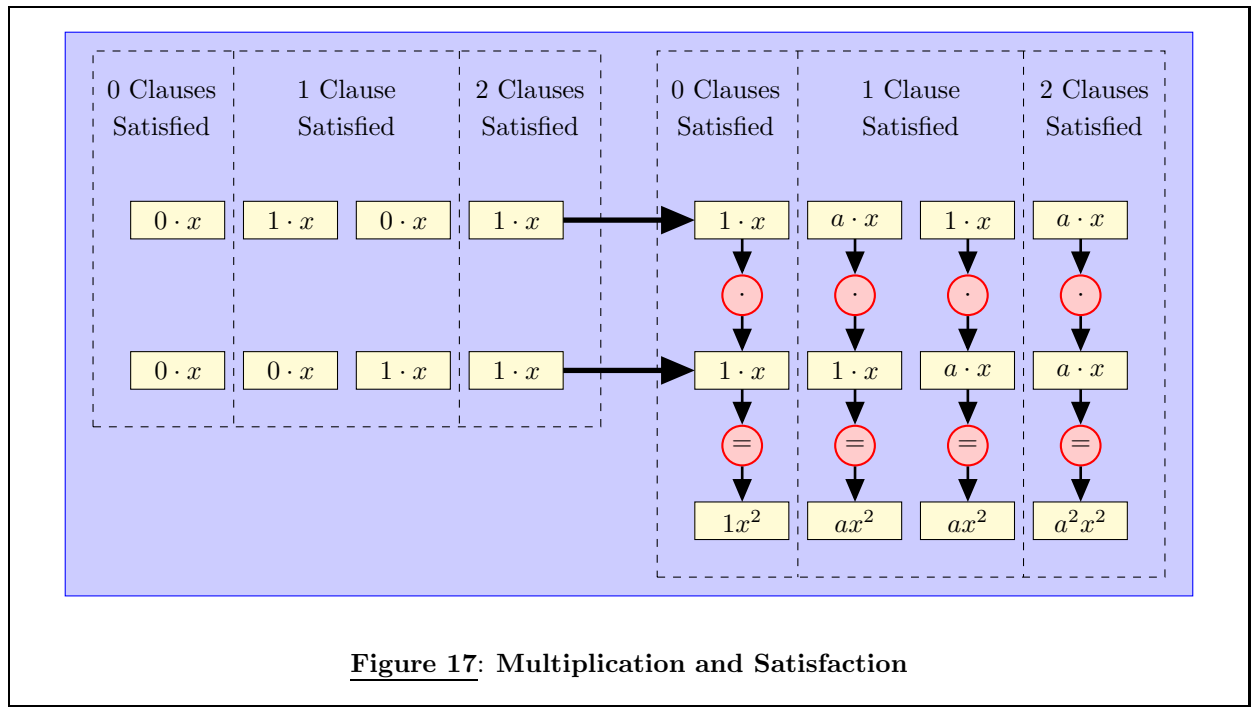
	\caption{\bf{Multiplication and Satisfaction}\label{MultiplicationSatisfaction}}
\end{PageFigure}

\pagebreak

\Subsection{Multiplication and\\ Satisfaction\label{MultiplicationSatisfactionSection}}
Figure \ref{MultiplicationSatisfaction} shows another viewpoint of arithmetic multiplication.  This time the four different cases based on the coefficients are shown.  In other words, between any two coefficients being multiplied, the input coefficients must be one of four cases.  Also, on the left, the satisfaction between the two original coefficients is shown.  Then, on the right, the corresponding modified coefficients and their results are shown.  Note that there are still really three cases of satisfaction, resulting in $1x^2$, $ax^2$, and $a^2x^2$.

It's seen that modifying the coefficients for multiplication has allowed for the three cases to appear in the results of multiplication.  This is one of the keys to getting things to work correctly.  The algorithm is really interested in the case when both clauses are satisfied.

Unfortunately, the results for multiplication mix the diagonal and off-diagonal cases together.  In order to isolate the diagonal, there will be some tricky interplay between multiplication and addition ahead.

\begin{ColumnFigure}
	\begin{center}
		\begin{tikzpicture}
			\filldraw [fill=blue!20, draw=blue] (-.5,.5) rectangle (6.25,-2.5);
			
			\foreach \x in {0,1,...,3}
			{
				\filldraw [fill=yellow!20, draw=black] (\x*1.5,0) rectangle (\x*1.5+1.25,-0.5);
				\filldraw [fill=yellow!20, draw=black] (\x*1.5,-0.5) rectangle (\x*1.5+1.25,-1.0);
				
				\filldraw [fill=yellow!20, draw=black] (\x*1.5,-1.5) rectangle (\x*1.5+1.25,-2.0);
			}
			
			\node at (.63 + 0*1.5,-.25) {$0\cdot x$};
			\node at (.63 + 1*1.5,-.25) {$1\cdot x$};
			\node at (.63 + 2*1.5,-.25) {$0\cdot x$};
			\node at (.63 + 3*1.5,-.25) {$1\cdot x$};
			
			\node at (.63 + 0*1.5,-.75) {$0\cdot x$};
			\node at (.63 + 1*1.5,-.75) {$0\cdot x$};
			\node at (.63 + 2*1.5,-.75) {$1\cdot x$};
			\node at (.63 + 3*1.5,-.75) {$1\cdot x$};
			
			\draw (0,-1.25) -- (5.75,-1.25);
			
			\node at (.63 + 0*1.5,-1.75) {$0\cdot x^2$};
			\node at (.63 + 1*1.5,-1.75) {$0\cdot x^2$};
			\node at (.63 + 2*1.5,-1.75) {$0\cdot x^2$};
			\node at (.63 + 3*1.5,-1.75) {$1\cdot x^2$};
			
		\end{tikzpicture}
	\end{center}
	\caption{\bf{Unmodified Multiplication}\label{UnmodifiedMultiplication}}
\end{ColumnFigure}

One thing that can be noted is that multiplication with the original clause polynomials isolates the case where everything is maximally satisfied (both clauses are satisfied).  This can be seen in Figure \ref{UnmodifiedMultiplication}.  Note that only the case on the far right (where both clauses have nonzero coefficients) returns anything other than zero.  Now this will occur for both diagonal and off-diagonal values, but it's very close to a solution.  The algorithm wants the diagonal portion of this result, and seeks to somehow eliminate the off-diagonal portion of it.

The next subsection will show how the algorithm can begin to seperate diagonal terms from off-diagonal terms.  The algorithm's efforts will be concentrated on seperating diagonal values from off-diagonal values, since the result will lead to a solution.

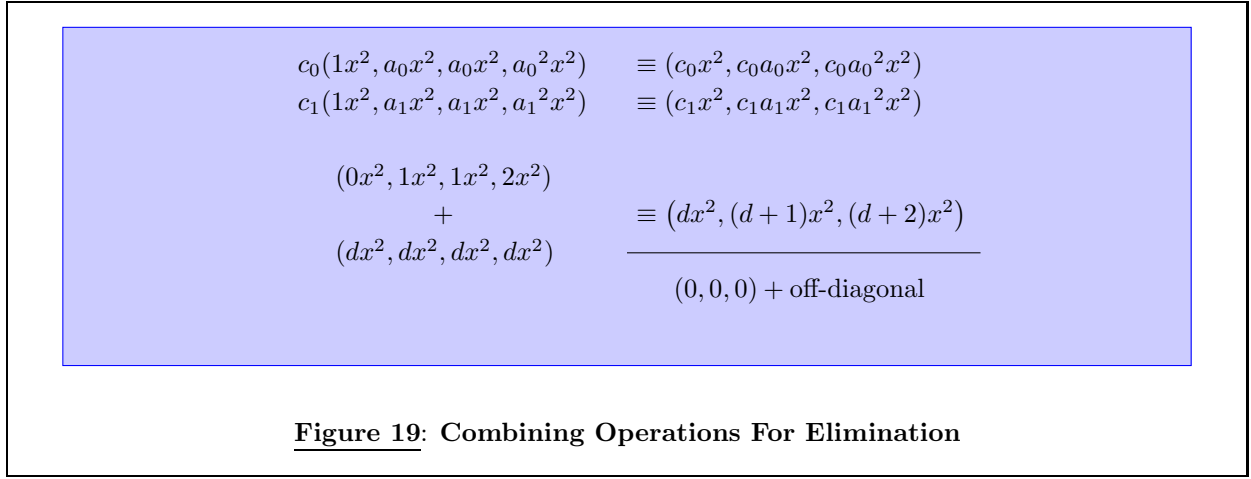
\begin{PageFigure}
	\begin{center}
		\begin{tikzpicture}
			\filldraw[fill=blue!20,draw=blue] (-3,.5) rectangle (12,-4.0);
		
			\node at (0  , 0  ) [right] {$c_0(1x^2, a_0x^2, a_0x^2, {a_0}^2x^2)$};
			\node at (0  ,-0.5) [right] {$c_1(1x^2, a_1x^2, a_1x^2, {a_1}^2x^2)$};
			
			\node at (0.5,-1.5) [right] {$(0x^2,1x^2,1x^2,2x^2)$};
			\node at (1.8,-2.0) [right] {+};
			\node at (0.5,-2.5) [right] {$(dx^2,dx^2,dx^2,dx^2)$};
			
			\node at (4.5, 0  ) [right] {$\equiv (c_0x^2,c_0a_0x^2,c_0{a_0}^2x^2)$};
			\node at (4.5,-0.5) [right] {$\equiv (c_1x^2,c_1a_1x^2,c_1{a_1}^2x^2)$};
			
			\node at (4.5,-2.0) [right] {$\equiv \left(dx^2,(d+1)x^2,(d+2)x^2\right)$};
			\draw (4.5,-2.5) -- (9.2,-2.5);
			\node at (5.0,-3.0) [right] {$(0,0,0) + \mbox{off-diagonal}$}; 			
		\end{tikzpicture}
	\end{center}
	\caption{\bf{Combining Operations For Elimination}\label{EliminationOperation}}
\end{PageFigure}

\pagebreak

\Subsection{Eliminating The Diagonal}
Figure \ref{Cases}, on the following page, presents the cases for addition alongside the better-explored operation of multiplication.  The main thing to note is that addition can return one of three results, just like the equivalence classes of satisfaction.  In fact, they are once again related in the same way that multiplication is related to satisfaction.

Figure \ref{DiagonalElimination} shows what is proposed for the algorithm.  It will take two multiplication operations (three are shown, but this is more than needed), and combine them with one addition, to eliminate everything except for the off-dagonal values, which are shown as gray triangles.

Everything is really summarized as arithmetic in Figure \ref{EliminationOperation}.  Here, the original results are shown on the left, being modified so that they can be combined together.  It can be noted that the middle two cases have always had equal results, so they have been combined. {\textit{The right side shows only three cases in paranthesis, which are the satisfaction equivalences.}}  The algorithm seeks to eliminate these, thus the sums at the bottom are zero, except for the off-diagonal.

Here the algorithm comes up with three equations that must be satisfied:

\begin{align}
c_0 + c_1 - d &= 0 \\
c_0a_0 + c_1a_1 - (d+1) &= 0 \\
c_0{a_0}^2 + c_1{a_1}^2 - (d+2) &= 0
\end{align}

Again, these come from the three cases on the right side of Figure \ref{EliminationOperation}, summing the columns.

To summarize all of this again, what is essentially happening is that two multiplications, together with an addition, cancel out all coefficients along the diagonal.  However, multiplication creates off-diagonal values, and these values will remain.  So in effect, the algorithm isolates off-diagonal values.  It does so by using the values calculated from addition to cancel out the diagonal from multiplication.

So now the algorithm can concentrate fully on eliminating the terms along the diagonal to isolate the off-diagonal terms.  The following subsection will explore how to get these terms to cancel.

First, it's time to introduce modular arithmetic.  Subsection \ref{ModularArithmeticBibliography} in the Appendix notes some sources that go over modular arithmetic.  To simplify the ideas, essentially all calculations are performed as usual, except that an additional operation is performed afterwards.  After the calculations are done, to get the result modulo a prime $p$, the algorithm does the equivalent of taking the remainder after dividing the result by $p$.  Thusly, all calculations are represented by an integer greater than or equal to zero and less than $p$.  So all calculations have a very limited range of values that can result.

This introduces a notion of equivalence, where two values are equivalent if they are the same modulo $p$.  This allows the algorithm to find solutions more easily, since the normal restriction that calculations must be equal is relaxed so that calculations must only be equivalent.

Equivalence is really introduced so that the equations that must be satisfied can be solved.  The relaxed restrictions allow for solutions that can be found easily.

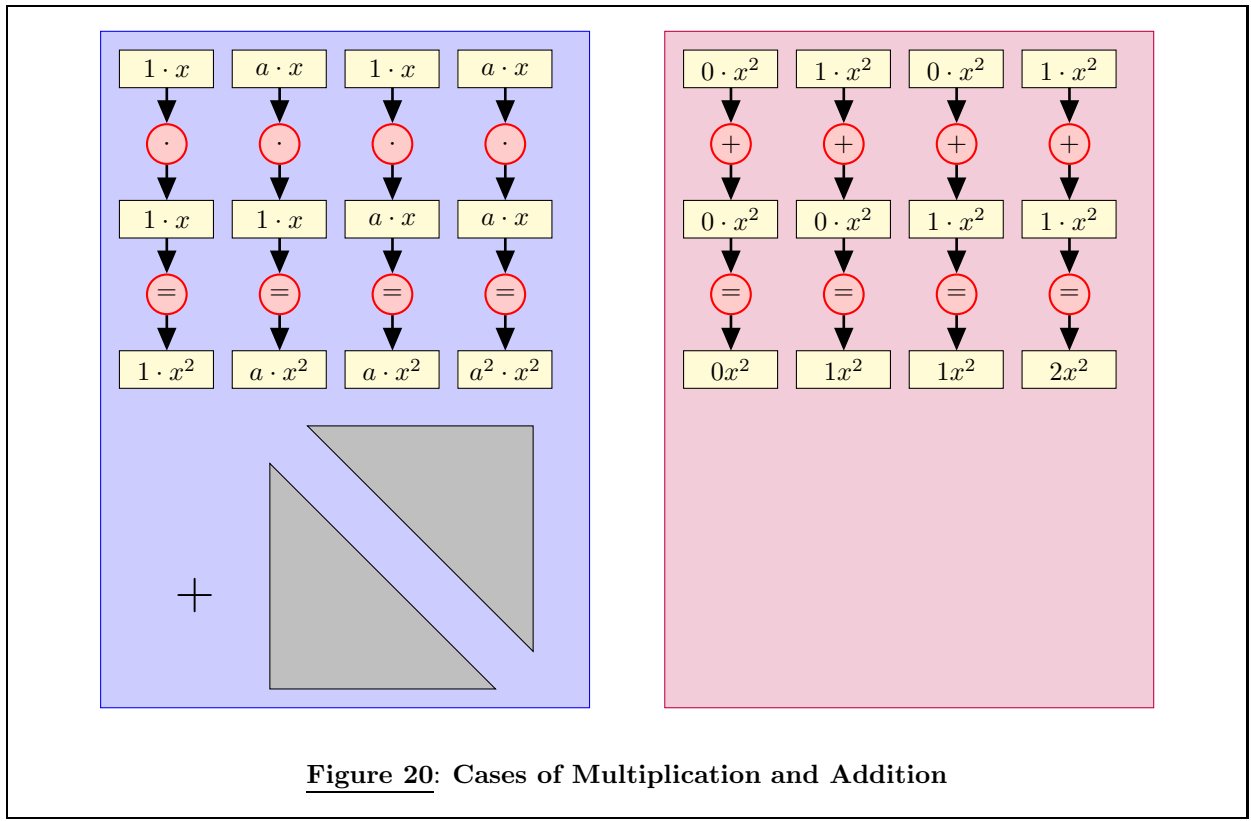
\begin{PageFigure}
	\begin{center}
		\begin{tikzpicture}
		
			\tikzstyle{Circ}=[circle,draw=red,fill=red!20,thick, inner sep=1pt,minimum size=15]
		
			\filldraw [fill=blue!20, draw=blue] (-.25,.25) rectangle (6.25,-8.75);
			\filldraw [fill=purple!20, draw=purple] (7.25,.25) rectangle (13.75,-8.75);

			\foreach \x in {0,1,...,3}
			{
				\filldraw [fill=yellow!20, draw=black] (\x*1.5,0) rectangle (\x*1.5+1.25,-0.5);
				
				\node (cdot\x) at (\x*1.5 + .63,-1.25) [Circ] {$\cdot$};
				\draw [line width = 1,-triangle 45] (\x*1.5 + .63,-0.5) -- (cdot\x);
				\draw [line width = 1,-triangle 45] (cdot\x) -- (\x*1.5 + .63,-2);
				\filldraw [fill=yellow!20, draw=black] (\x*1.5,-2) rectangle (\x*1.5+1.25,-2.5);
				
				\node (cdot2\x) at (\x*1.5 + .63,-3.25) [Circ] {=};
				\draw [line width = 1,-triangle 45] (\x*1.5 + .63,-2.5) -- (cdot2\x);
				\draw [line width = 1,-triangle 45] (cdot2\x) -- (\x*1.5 + .63,-4);
				\filldraw [fill=yellow!20, draw=black] (\x*1.5,-4) rectangle (\x*1.5+1.25,-4.5);
			}
			
			\foreach \x in {5,6,...,8}
			{
				\filldraw [fill=yellow!20, draw=black] (\x*1.5,0) rectangle (\x*1.5+1.25,-0.5);
				
				\node (cdot\x) at (\x*1.5 + .63,-1.25) [Circ] {+};
				\draw [line width = 1,-triangle 45] (\x*1.5 + .63,-0.5) -- (cdot\x);
				\draw [line width = 1,-triangle 45] (cdot\x) -- (\x*1.5 + .63,-2);
				\filldraw [fill=yellow!20, draw=black] (\x*1.5,-2) rectangle (\x*1.5+1.25,-2.5);
				
				\node (cdot2\x) at (\x*1.5 + .63,-3.25) [Circ] {=};
				\draw [line width = 1,-triangle 45] (\x*1.5 + .63,-2.5) -- (cdot2\x);
				\draw [line width = 1,-triangle 45] (cdot2\x) -- (\x*1.5 + .63,-4);
				\filldraw [fill=yellow!20, draw=black] (\x*1.5,-4) rectangle (\x*1.5+1.25,-4.5);
			}
			
			\node (1topmult) at (.63 + 0*1.5,-.25) {$1\cdot x$};
			\node (2topmult) at (.63 + 1*1.5,-.25) {$a\cdot x$};
			\node (3topmult) at (.63 + 2*1.5,-.25) {$1\cdot x$};
			\node (4topmult) at (.63 + 3*1.5,-.25) {$a\cdot x$};
			
			\node (1botmult) at (.63 + 0*1.5,-2.25) {$1\cdot x$};
			\node (2botmult) at (.63 + 1*1.5,-2.25) {$1\cdot x$};
			\node (3botmult) at (.63 + 2*1.5,-2.25) {$a\cdot x$};
			\node (4botmult) at (.63 + 3*1.5,-2.25) {$a\cdot x$};
			
			\node (1resmult) at (.63 + 0*1.5,-4.25) {$1\cdot x^2$};
			\node (2resmult) at (.63 + 1*1.5,-4.25) {$a\cdot x^2$};
			\node (3resmult) at (.63 + 2*1.5,-4.25) {$a\cdot x^2$};
			\node (4resmult) at (.63 + 3*1.5,-4.25) {$a^2\cdot x^2$};

			\node (1topadd) at (.63 + 5*1.5,-.25) {$0\cdot x^2$};
			\node (2topadd) at (.63 + 6*1.5,-.25) {$1\cdot x^2$};
			\node (3topadd) at (.63 + 7*1.5,-.25) {$0\cdot x^2$};
			\node (4topadd) at (.63 + 8*1.5,-.25) {$1\cdot x^2$};
			
			\node (1botadd) at (.63 + 5*1.5,-2.25) {$0\cdot x^2$};
			\node (2botadd) at (.63 + 6*1.5,-2.25) {$0\cdot x^2$};
			\node (3botadd) at (.63 + 7*1.5,-2.25) {$1\cdot x^2$};
			\node (4botadd) at (.63 + 8*1.5,-2.25) {$1\cdot x^2$};
			
			\node (1resadd) at (.63 + 5*1.5,-4.25) {$0x^2$};
			\node (2resadd) at (.63 + 6*1.5,-4.25) {$1x^2$};
			\node (3resadd) at (.63 + 7*1.5,-4.25) {$1x^2$};
			\node (4resadd) at (.63 + 8*1.5,-4.25) {$2x^2$};
			
			\filldraw[fill=gray!50, draw=black] (2,-5.5) -- (2,-8.5) -- (5,-8.5) -- cycle; 
			\filldraw[fill=gray!50, draw=black] (2.5,-5) -- (5.5,-5) -- (5.5,-8) -- cycle;
			
			\node at (1,-7.25) {\huge{+}};
			
		\end{tikzpicture}
	\end{center}
	\caption{\bf{Cases of Multiplication and Addition}\label{Cases}}
\end{PageFigure}

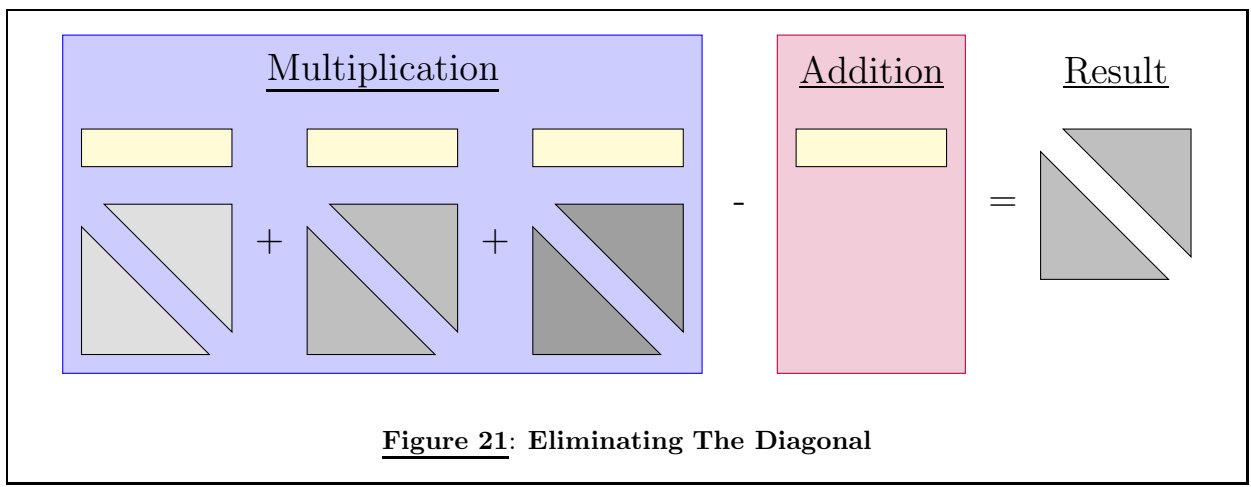
\begin{PageFigure}
	\begin{center}
		\begin{tikzpicture}
				\filldraw [fill=blue!20,draw=blue] (-.25,1.25) rectangle (8.25,-3.25);
		\foreach \shade in {25,50,75}
		{
			\filldraw [fill=yellow!20] (\shade*0.12-3,0) rectangle (\shade*0.12-1,-0.5);
			\filldraw [fill=gray!\shade] (\shade*0.12-3+0.3,-1) -- (\shade*0.12-1,-1) --
				(\shade*0.12-1,-2.7) -- cycle;
			\filldraw [fill=gray!\shade] (\shade*0.12-3,-1.3) -- (\shade*0.12-1-.3,-3) --
				(\shade*0.12-3,-3) -- cycle;
		}
		
		\node at (2.5, -1.5) {\Large{+}};
		\node at (5.5, -1.5) {\Large{+}};
		\node at (4,.75) {$\underline{\mbox{\Large{Multiplication}}}$};
		
		\node at (8.75,-1) {\Large{-}};
		
		\filldraw [fill=purple!20,draw=purple] (9.25,1.25) rectangle (11.75,-3.25);
		\filldraw [fill=yellow!20] (9.5,0) rectangle (11.5,-.5);
		\node at (10.5,.75) {$\underline{\mbox{\Large{Addition}}}$};
		
		\node at (12.25,-1) {\Large{=}};
		
		\filldraw [fill=gray!50] (13.05,0) -- (14.75,0) -- (14.75,-1.7) -- cycle;
		\filldraw [fill=gray!50] (12.75,-0.3) -- (14.45,-2) -- (12.75,-2) -- cycle;
		\node at (13.75,.75) {$\underline{\mbox{\Large{Result}}}$};
		
			
		\end{tikzpicture}
	\end{center}
	\caption{\bf{Eliminating The Diagonal}\label{DiagonalElimination}}
\end{PageFigure}

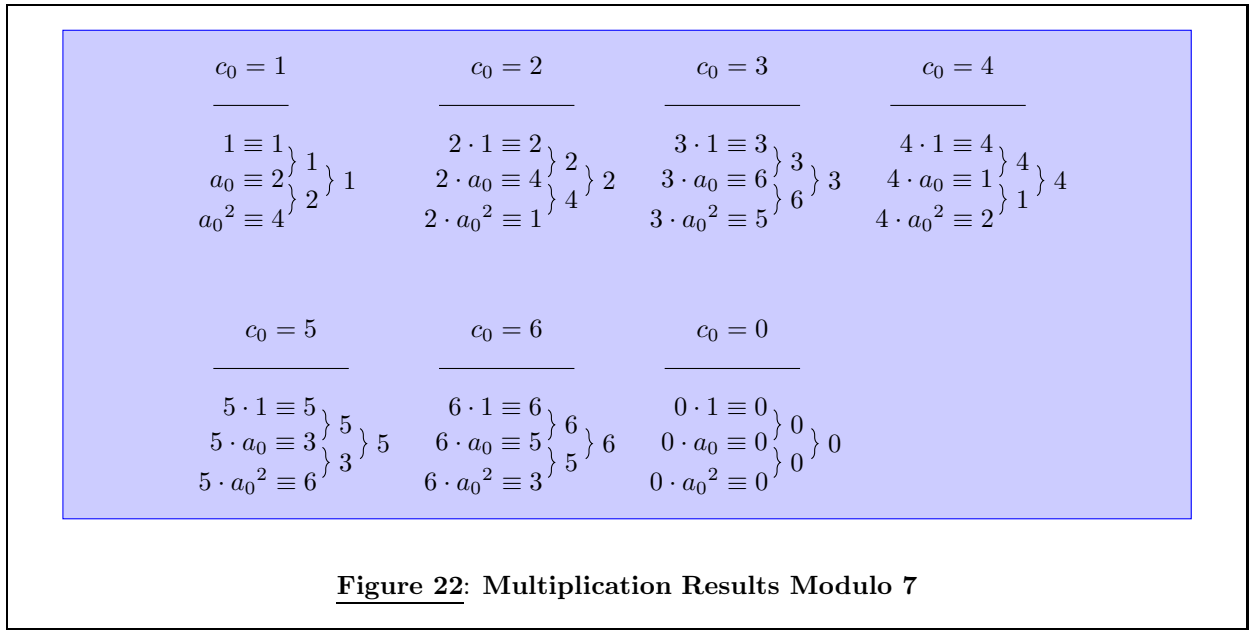
\begin{PageFigure}
	\begin{center}
		\begin{tikzpicture}
			\filldraw[fill=blue!20, draw=blue] (-2,1.5) rectangle (13,-5);
		
			\node at (0.5,1) {$c_0 = 1$};
			\node at (3.9,1) {$c_0 = 2$};
			\node at (6.9,1) {$c_0 = 3$};
			\node at (9.9,1) {$c_0 = 4$};
			\node at (0.9,-2.5) {$c_0 = 5$};
			\node at (3.9,-2.5) {$c_0 = 6$};
			\node at (6.9,-2.5) {$c_0 = 0$};
			
			\draw (0,0.5) -- (1,0.5);
			\draw (3,0.5) -- (4.8,0.5);
			\draw (6,0.5) -- (7.8,0.5);
			\draw (9,0.5) -- (10.8,0.5);
			\draw (0,-3) -- (1.8,-3);
			\draw (3,-3) -- (4.8,-3);
			\draw (6,-3) -- (7.8,-3);
		
			\node at ( 0   , 0  ) [right] {$1 \equiv 1$};
			\node at (-0.17,-0.5) [right] {$a_0 \equiv 2$};
			\node at (-0.32,-1.0) [right] {${a_0}^2 \equiv 4$};
			\draw [snake=brace] (1.0,-0.05) -- (1.0,-0.45);
			\draw [snake=brace] (1.0,-0.55) -- (1.0,-0.95);
			\node at (1.1,-0.25) [right] {$1$};
			\node at (1.1,-0.75) [right] {$2$};
			\draw [snake=brace] (1.5,-0.30) -- (1.5,-0.70);
			\node at (1.6,-0.50) [right] {$1$};
			
			\node at ( 0   +3, 0  ) [right] {$2\cdot 1 \equiv 2$};
			\node at (-0.17+3,-0.5) [right] {$2\cdot a_0 \equiv 4$};
			\node at (-0.32+3,-1.0) [right] {$2\cdot {a_0}^2 \equiv 1$};
			\draw [snake=brace] (1.45+3,-0.05) -- (1.45+3,-0.45);
			\draw [snake=brace] (1.45+3,-0.55) -- (1.45+3,-0.95);
			\node at (1.55+3,-0.25) [right] {$2$};
			\node at (1.55+3,-0.75) [right] {$4$};
			\draw [snake=brace] (1.95+3,-0.30) -- (1.95+3,-0.70);
			\node at (2.05+3,-0.50) [right] {$2$};
			
			\node at ( 0   +6, 0  ) [right] {$3\cdot 1 \equiv 3$};
			\node at (-0.17+6,-0.5) [right] {$3\cdot a_0 \equiv 6$};
			\node at (-0.32+6,-1.0) [right] {$3\cdot {a_0}^2 \equiv 5$};
			\draw [snake=brace] (1.45+6,-0.05) -- (1.45+6,-0.45);
			\draw [snake=brace] (1.45+6,-0.55) -- (1.45+6,-0.95);
			\node at (1.55+6,-0.25) [right] {$3$};
			\node at (1.55+6,-0.75) [right] {$6$};
			\draw [snake=brace] (1.95+6,-0.30) -- (1.95+6,-0.70);
			\node at (2.05+6,-0.50) [right] {$3$};
			
			\node at ( 0   +9, 0  ) [right] {$4\cdot 1 \equiv 4$};
			\node at (-0.17+9,-0.5) [right] {$4\cdot a_0 \equiv 1$};
			\node at (-0.32+9,-1.0) [right] {$4\cdot {a_0}^2 \equiv 2$};
			\draw [snake=brace] (1.45+9,-0.05) -- (1.45+9,-0.45);
			\draw [snake=brace] (1.45+9,-0.55) -- (1.45+9,-0.95);
			\node at (1.55+9,-0.25) [right] {$4$};
			\node at (1.55+9,-0.75) [right] {$1$};
			\draw [snake=brace] (1.95+9,-0.30) -- (1.95+9,-0.70);
			\node at (2.05+9,-0.50) [right] {$4$};

			\node at ( 0   +0,-3.5) [right] {$5\cdot 1 \equiv 5$};
			\node at (-0.17+0,-4.0) [right] {$5\cdot a_0 \equiv 3$};
			\node at (-0.32+0,-4.5) [right] {$5\cdot {a_0}^2 \equiv 6$};
			\draw [snake=brace] (1.45+0,-3.55) -- (1.45+0,-3.95);
			\draw [snake=brace] (1.45+0,-4.05) -- (1.45+0,-4.45);
			\node at (1.55+0,-3.75) [right] {$5$};
			\node at (1.55+0,-4.25) [right] {$3$};
			\draw [snake=brace] (1.95+0,-3.80) -- (1.95+0,-4.20);
			\node at (2.05+0,-4.00) [right] {$5$};
			
			\node at ( 0   +3,-3.5) [right] {$6\cdot 1 \equiv 6$};
			\node at (-0.17+3,-4.0) [right] {$6\cdot a_0 \equiv 5$};
			\node at (-0.32+3,-4.5) [right] {$6\cdot {a_0}^2 \equiv 3$};
			\draw [snake=brace] (1.45+3,-3.55) -- (1.45+3,-3.95);
			\draw [snake=brace] (1.45+3,-4.05) -- (1.45+3,-4.45);
			\node at (1.55+3,-3.75) [right] {$6$};
			\node at (1.55+3,-4.25) [right] {$5$};
			\draw [snake=brace] (1.95+3,-3.80) -- (1.95+3,-4.20);
			\node at (2.05+3,-4.00) [right] {$6$};
			
			\node at ( 0   +6,-3.5) [right] {$0\cdot 1 \equiv 0$};
			\node at (-0.17+6,-4.0) [right] {$0\cdot a_0 \equiv 0$};
			\node at (-0.32+6,-4.5) [right] {$0\cdot {a_0}^2 \equiv 0$};
			\draw [snake=brace] (1.45+6,-3.55) -- (1.45+6,-3.95);
			\draw [snake=brace] (1.45+6,-4.05) -- (1.45+6,-4.45);
			\node at (1.55+6,-3.75) [right] {$0$};
			\node at (1.55+6,-4.25) [right] {$0$};
			\draw [snake=brace] (1.95+6,-3.80) -- (1.95+6,-4.20);
			\node at (2.05+6,-4.00) [right] {$0$};	
			
		\end{tikzpicture}
	\end{center}
	\caption{\bf{Multiplication Results Modulo 7}\label{MultiplicationModulo7}}
\end{PageFigure}
\pagebreak

To find solutions that eliminate the diagonal, a finite field is introduced, allowing operations to be conducted modulo a prime $p$.  Figure \ref{MultiplicationModulo7} displays calculations inside a field modulo 7.  Here the algorithm selects $a_0 = 2$, although it could select really any value other than one or zero.  It then calculates the results of multiplication (times a constant $c_0$), which for the various equivalence classes are $c_0 \cdot 1$, $c_0 \cdot a_0$, and $c_0 \cdot {a_0}^2$.  

The algorithm is really interested in the second-order difference.  This is illustrated in Figure \ref{MultiplicationDifferences}.  That is, it takes the difference between results, and then takes the difference of these differences.

It's fairly straightfoward to get the initial results; using arithmetic multiplication works fine.  Then the results have to be modulated.  As an example, in Figure \ref{MultiplicationDifferences}, $a_1$ is set as three.  Then, to calculate ${a_1}^2$, take $3^2=9$.  Then $9 / 2 = 1$ with remainder 2.  So the result is the remainder, which is 2.

That gets the initial results, in the field.  Then to get the first order differences, take successive results and subtract the first from the successive.  This is illustrated very clearly in the figure, which does a better job of explaining.  The take the difference again, which is called the second-order difference.

\begin{ColumnFigure}
	\begin{center}
		\begin{tikzpicture}
			\filldraw [fill=blue!20, draw=blue] (-.5,.5) rectangle (6.25,-1.5);
			
			\node at ( 0   , 0  ) [right] {$1 \equiv 1 \bmod 7$};
			\node at (-0.17,-0.5) [right] {$a_1 \equiv 3 \bmod 7$};
			\node at (-0.32,-1.0) [right] {${a_1}^2 \equiv 2 \bmod 7$};
			\draw [snake=brace] (2.15,-0.05) -- (2.15,-0.45);
			\draw [snake=brace] (2.15,-0.55) -- (2.15,-0.95);
			\node at (2.25,-0.25) [right] {$3-1 \equiv 2$};
			\node at (2.25,-0.75) [right] {$2-3 \equiv 6$};
			\draw [snake=brace] (4.0,-0.30) -- (4.0,-0.70);
			\node at (4.1,-0.50) [right] {$6-2 \equiv 4$};
			
		\end{tikzpicture}
	\end{center}
	\caption{\bf{Differences of Multiplication Results}\label{MultiplicationDifferences}}
\end{ColumnFigure}



The reason that it's so important to get these second-order differences is that they help match up multiplication results.  Note that both figures are actually two seperate multiplications; Figure \ref{MultiplicationModulo7} uses $a_0$ and $c_0$ while Figure \ref{MultiplicationDifferences} uses $a_1$ and $c_1$.  They both use the same prime, so they are two seperate multiplication results that can be matched up.  The goal is to find the second-order differences that add up to zero mod $p$.  Note that the example with $a_1$ has a second-order difference of four.  $3 + 4 = 7 \equiv 0 \bmod 7$, so the idea is to find a second-order difference of three in the other equation.  It is done with $c_0 = 3$.  So now two equations have been found that match up.

To check this result, the equations can be combined:

\begin{align}
\nonumber c_0 \cdot 1 + c_1 \cdot 1 &\equiv 3\cdot 1 + 1\cdot1 &\equiv 4 &\equiv d+0e \\
\nonumber c_0 \cdot a_0 + c_1 \cdot a_1 &\equiv 3\cdot 2 + 1\cdot3 &\equiv 2 &\equiv d+1e \\
\nonumber c_0 \cdot {a_0}^2 + c_1 \cdot {a_1}^2 &\equiv 3\cdot 4 + 1\cdot2 &\equiv 0 &\equiv d+2e
\end{align}

The main thing to note is that the sequence of results in the equations (4, 2, 0) can be recreated by an addition operation (on clause polynomials).  That is, the first equation $=4+0(-2)$, the second $=4+1(-2)$, and the third $=4+2(-2)$.

\begin{PageFigure}
	\begin{center}
		\begin{tikzpicture}
			\filldraw[fill=blue!20,draw=blue] (-1.35,1.25) rectangle (13.65,-2.25);
			\node at (6.15,0.75) {\Large{Diagonal}};
		
			\node at (0  , 0  ) [right] {$(c_0 \cdot 1) + (c_1 \cdot 1) - d$};
			\node at (0  ,-0.5) [right] {$(c_0 \cdot a_0) + (c_1 \cdot a_1) - (d-e)$};
			\node at (0  ,-1  ) [right] {$(c_0 \cdot {a_0}^2) + (c_1 \cdot {a_1}^2) - (d-2e)$};
			\node at (4.8, 0  ) [right] {$= 0$};
			\node at (4.8,-0.5) [right] {$= 0$};
			\node at (4.8,-1  ) [right] {$= 0$};
			\draw (0,-1.5) -- (5.6,-1.5);
			\node at (5.17,-2  ) [right] {$0$};
			
			\node at ( 7  , 0  ) [right] {$(3 \cdot 1) + (1 \cdot 1) - 4$};
			\node at ( 7  ,-0.5) [right] {$(3 \cdot 2) + (1 \cdot 3) - (4-2)$};
			\node at ( 7  ,-1  ) [right] {$(3 \cdot 2^2) + (1 \cdot 3^2) - (4-2\cdot2)$};
			\node at (11.5, 0  ) [right] {$= 0$};
			\node at (11.5,-0.5) [right] {$= 0$};
			\node at (11.5,-1  ) [right] {$= 0$};
			\draw (7,-1.5) -- (12.3,-1.5);
			\node at (11.86,-2  ) [right] {$0$};

			\filldraw[fill=purple!20,draw=purple] (-1.35,-2.75) rectangle (13.65,-6.25);
			\node at (6.15,-3.25) {\Large{Off-Diagonal}};
					
			\node at (0  ,-4  ) [right] {$(c_0 \cdot 1) + (c_1 \cdot 1)$};
			\node at (0  ,-4.5) [right] {$(c_0 \cdot a_0) + (c_1 \cdot a_1)$};
			\node at (0  ,-5  ) [right] {$(c_0 \cdot {a_0}^2) + (c_1 \cdot {a_1}^2)$};
			
			\node at ( 7  ,-4  ) [right] {$(3 \cdot 1) + (1 \cdot 1)$};
			\node at ( 7  ,-4.5) [right] {$(3 \cdot 2) + (1 \cdot 3)$};
			\node at ( 7  ,-5  ) [right] {$(3 \cdot 2^2) + (1 \cdot 3^2)$};
			\node at (11.5,-4  ) [right] {$= 4$};
			\node at (11.5,-4.5) [right] {$= 2$};
			\node at (11.5,-5  ) [right] {$= 0$};
			\draw (7,-5.5) -- (12.3,-5.5);
			\node at (9.65,-6  ) {$4b_0 + 2b_1 + 0b_2$};
			
		\end{tikzpicture}
	\end{center}
	\caption{\bf{Two Clause Results}\label{TwoClauseEquation}}
\end{PageFigure}
\pagebreak

\pagebreak
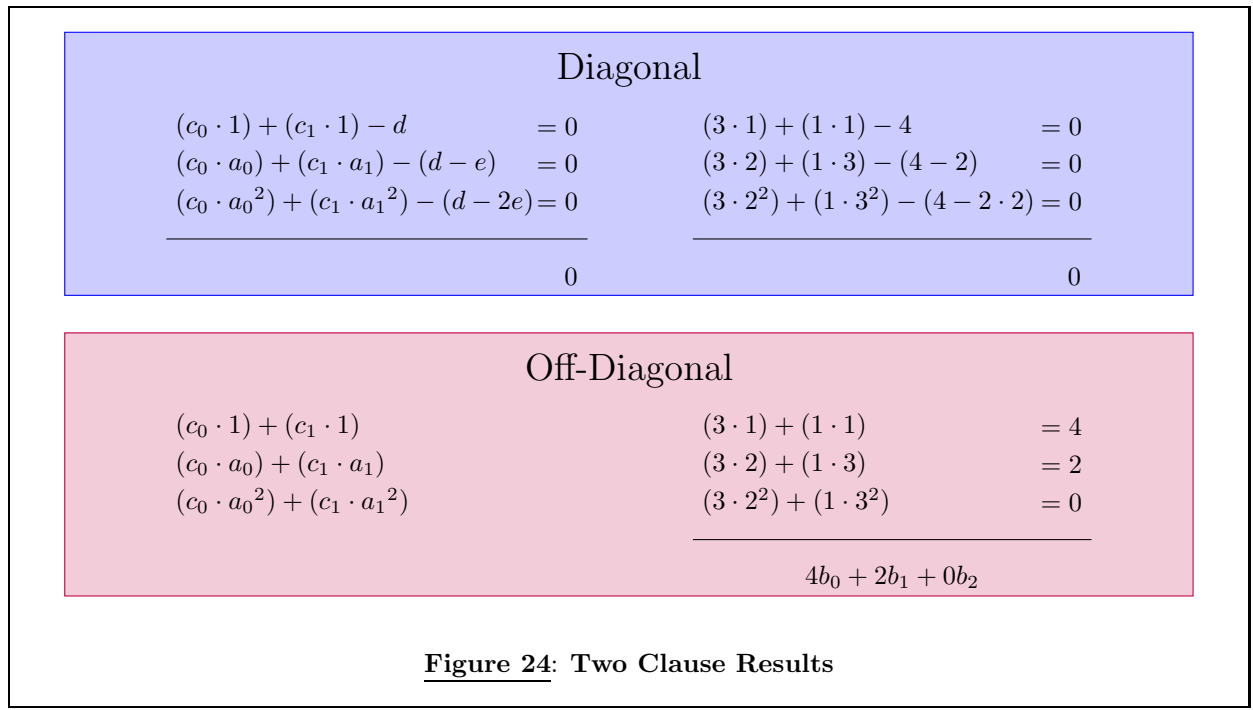
Figure \ref{TwoClauseEquation} shows the results of combining equations (addition and multiplication).  Here the the results of clause operations (multiplication and addition) are combined, and are shown according to whether or not they lie along the diagonal.  As mentioned previously, the addition operation cancels out the multiplication along the diagonal.  However, the results off the diagonal are not cancelled.

Thusly, the results off the diagonal become multipliers for unknown quantities.  This is because the polynomials that underline the equations are also part of this mix, and the resulting polynomial quantities are unknown, even though the multipliers are known.  Thus these quantities are represented as $b_0$, $b_1$, and $b_2$.

This allows for a new result.  Remember that the original quantities are combined together in the previous equations that were used.  Thus, this new result represents a sum of the new quantities, along with their respective multipliers.  Therefore, the algorithm arrives at a new result which is a single equation in three unknowns.

Again, time for some ideas.  Back in Section \ref{MultiplicationSatisfactionSection} on page \pageref{MultiplicationSatisfactionSection}, Figure \ref{UnmodifiedMultiplication} is discussed.  Especially important is the fact that the maximally satisfied clauses, both along the diagonal and off the diagonal, can be isolated.  In this section it has been shown that equations for off-diagonal values can be obtained.  Thinking about off-diagonal values, it is possible to create other equations using multiplication and addition - and these equations can help determine the off-diagonal values by combining them with the first equation and then using linear algebra to determine the values.

So essentially, the algorithm will use multiplication and addition of modified clause polynomials to create equations in three unknowns.  These unknowns are the off-diagonal values.  Then, the equations created can be combined via linear algebra to detetmine the off-diagonal values.  As mentioned previously, the algorithm can already isolate the maximally satisfied clause values together.  Unfortunately, the off-diagonal and diagonal values are combined together.  However, since the off-diagonal portion can be determined via linear algebra, the diagonal portion can be isolated from the off-diagonal portion using simple algebra, and the results from linear algebra (which give the off-diagonal portion).

The diagonal value of maximally satisfied clauses is thus determined, and this corresponds to the number of satisfied solutions.  Now if this number is nonzero, the whole problem can be satisfied.  Otherwise, the original problem can't be satisfied.

\pagebreak

\begin{PageFigure}
	\begin{center}
		\begin{tikzpicture}
			\filldraw[fill=blue!20, draw=blue] (-2,1.5) rectangle (13,-12);
		
			\node at (0.5,1) {$c_0 = 1$};
			\node at (3.9,1) {$c_0 = 2$};
			\node at (6.9,1) {$c_0 = 3$};
			\node at (9.9,1) {$c_0 = 4$};
			\node at (0.9,-2.5) {$c_0 = 5$};
			\node at (3.9,-2.5) {$c_0 = 6$};
			\node at (6.9,-2.5) {$c_0 = 7$};
			\node at (9.9,-2.5) {$c_0 = 8$};
			\node at (0.9,-6) {$c_0 = 9$};
			\node at (3.9,-6) {$c_0 = 10$};
			\node at (6.9,-6) {$c_0 = 11$};
			\node at (9.9,-6) {$c_0 = 12$};
			\node at (0.9,-9.5) {$c_0 = 13$};
			\node at (3.9,-9.5) {$c_0 = 14$};
			\node at (6.9,-9.5) {$c_0 = 15$};
			\node at (9.9,-9.5) {$c_0 = 16$};
			
			\draw (0,0.5) -- (1,0.5);
			\draw (3,0.5) -- (4.8,0.5);
			\draw (6,0.5) -- (7.8,0.5);
			\draw (9,0.5) -- (10.8,0.5);
			\draw (0,-3) -- (1.8,-3);
			\draw (3,-3) -- (4.8,-3);
			\draw (6,-3) -- (7.8,-3);
			\draw (9,-3) -- (10.8,-3);
			\draw (0,-6.5) -- (1.8,-6.5);
			\draw (3,-6.5) -- (4.8,-6.5);
			\draw (6,-6.5) -- (7.8,-6.5);
			\draw (9,-6.5) -- (10.8,-6.5);
			\draw (0,-10) -- (1.8,-10);
			\draw (3,-10) -- (4.8,-10);
			\draw (6,-10) -- (7.8,-10);
			\draw (9,-10) -- (10.8,-10);
		
			\node at ( 0   , 0  ) [right] {$1 \equiv 1$};
			\node at (-0.17,-0.5) [right] {$a_0 \equiv 2$};
			\node at (-0.32,-1.0) [right] {${a_0}^2 \equiv 4$};
			\draw [snake=brace] (1.0,-0.05) -- (1.0,-0.45);
			\draw [snake=brace] (1.0,-0.55) -- (1.0,-0.95);
			\node at (1.1,-0.25) [right] {$1$};
			\node at (1.1,-0.75) [right] {$2$};
			\draw [snake=brace] (1.5,-0.30) -- (1.5,-0.70);
			\node at (1.6,-0.50) [right] {$1$};
			
			\node at ( 0   +3, 0  ) [right] {$2\cdot 1 \equiv 2$};
			\node at (-0.17+3,-0.5) [right] {$2\cdot a_0 \equiv 4$};
			\node at (-0.32+3,-1.0) [right] {$2\cdot {a_0}^2 \equiv 8$};
			\draw [snake=brace] (1.45+3,-0.05) -- (1.45+3,-0.45);
			\draw [snake=brace] (1.45+3,-0.55) -- (1.45+3,-0.95);
			\node at (1.55+3,-0.25) [right] {$2$};
			\node at (1.55+3,-0.75) [right] {$4$};
			\draw [snake=brace] (1.95+3,-0.30) -- (1.95+3,-0.70);
			\node at (2.05+3,-0.50) [right] {$2$};
			
			\node at ( 0   +6, 0  ) [right] {$3\cdot 1 \equiv 3$};
			\node at (-0.17+6,-0.5) [right] {$3\cdot a_0 \equiv 6$};
			\node at (-0.32+6,-1.0) [right] {$3\cdot {a_0}^2 \equiv 12$};
			\draw [snake=brace] (1.45+6,-0.05) -- (1.45+6,-0.45);
			\draw [snake=brace] (1.45+6,-0.55) -- (1.45+6,-0.95);
			\node at (1.55+6,-0.25) [right] {$3$};
			\node at (1.55+6,-0.75) [right] {$6$};
			\draw [snake=brace] (1.95+6,-0.30) -- (1.95+6,-0.70);
			\node at (2.05+6,-0.50) [right] {$3$};
			
			\node at ( 0   +9, 0  ) [right] {$4\cdot 1 \equiv 4$};
			\node at (-0.17+9,-0.5) [right] {$4\cdot a_0 \equiv 8$};
			\node at (-0.32+9,-1.0) [right] {$4\cdot {a_0}^2 \equiv 16$};
			\draw [snake=brace] (1.45+9,-0.05) -- (1.45+9,-0.45);
			\draw [snake=brace] (1.45+9,-0.55) -- (1.45+9,-0.95);
			\node at (1.55+9,-0.25) [right] {$4$};
			\node at (1.55+9,-0.75) [right] {$8$};
			\draw [snake=brace] (1.95+9,-0.30) -- (1.95+9,-0.70);
			\node at (2.05+9,-0.50) [right] {$4$};

			\node at ( 0   +0,-3.5) [right] {$5\cdot 1 \equiv 5$};
			\node at (-0.17+0,-4.0) [right] {$5\cdot a_0 \equiv 10$};
			\node at (-0.32+0,-4.5) [right] {$5\cdot {a_0}^2 \equiv 3$};
			\draw [snake=brace] (1.45+0,-3.55) -- (1.45+0,-3.95);
			\draw [snake=brace] (1.45+0,-4.05) -- (1.45+0,-4.45);
			\node at (1.55+0,-3.75) [right] {$5$};
			\node at (1.55+0,-4.25) [right] {$10$};
			\draw [snake=brace] (1.95+0,-3.80) -- (1.95+0,-4.20);
			\node at (2.05+0,-4.00) [right] {$5$};
			
			\node at ( 0   +3,-3.5) [right] {$6\cdot 1 \equiv 6$};
			\node at (-0.17+3,-4.0) [right] {$6\cdot a_0 \equiv 12$};
			\node at (-0.32+3,-4.5) [right] {$6\cdot {a_0}^2 \equiv 7$};
			\draw [snake=brace] (1.45+3,-3.55) -- (1.45+3,-3.95);
			\draw [snake=brace] (1.45+3,-4.05) -- (1.45+3,-4.45);
			\node at (1.55+3,-3.75) [right] {$6$};
			\node at (1.55+3,-4.25) [right] {$12$};
			\draw [snake=brace] (1.95+3,-3.80) -- (1.95+3,-4.20);
			\node at (2.05+3,-4.00) [right] {$6$};
			
			\node at ( 0   +6,-3.5) [right] {$7\cdot 1 \equiv 7$};
			\node at (-0.17+6,-4.0) [right] {$7\cdot a_0 \equiv 14$};
			\node at (-0.32+6,-4.5) [right] {$7\cdot {a_0}^2 \equiv 11$};
			\draw [snake=brace] (1.45+6,-3.55) -- (1.45+6,-3.95);
			\draw [snake=brace] (1.45+6,-4.05) -- (1.45+6,-4.45);
			\node at (1.55+6,-3.75) [right] {$7$};
			\node at (1.55+6,-4.25) [right] {$14$};
			\draw [snake=brace] (1.95+6,-3.80) -- (1.95+6,-4.20);
			\node at (2.05+6,-4.00) [right] {$7$};	
			
			\node at ( 0   +9,-3.5) [right] {$7\cdot 1 \equiv 8$};
			\node at (-0.17+9,-4.0) [right] {$7\cdot a_0 \equiv 16$};
			\node at (-0.32+9,-4.5) [right] {$7\cdot {a_0}^2 \equiv 15$};
			\draw [snake=brace] (1.45+9,-3.55) -- (1.45+9,-3.95);
			\draw [snake=brace] (1.45+9,-4.05) -- (1.45+9,-4.45);
			\node at (1.55+9,-3.75) [right] {$8$};
			\node at (1.55+9,-4.25) [right] {$16$};
			\draw [snake=brace] (1.95+9,-3.80) -- (1.95+9,-4.20);
			\node at (2.05+9,-4.00) [right] {$8$};

			\node at ( 0   +0,-7.0) [right] {$5\cdot 1 \equiv 9$};
			\node at (-0.17+0,-7.5) [right] {$5\cdot a_0 \equiv 1$};
			\node at (-0.32+0,-8.0) [right] {$5\cdot {a_0}^2 \equiv 2$};
			\draw [snake=brace] (1.45+0,-7.05) -- (1.45+0,-7.45);
			\draw [snake=brace] (1.45+0,-7.55) -- (1.45+0,-7.95);
			\node at (1.55+0,-7.25) [right] {$9$};
			\node at (1.55+0,-7.75) [right] {$1$};
			\draw [snake=brace] (1.95+0,-7.30) -- (1.95+0,-7.70);
			\node at (2.05+0,-7.50) [right] {$9$};
			
			\node at ( 0   +3,-7.0) [right] {$5\cdot 1 \equiv 10$};
			\node at (-0.17+3,-7.5) [right] {$5\cdot a_0 \equiv 3$};
			\node at (-0.32+3,-8.0) [right] {$5\cdot {a_0}^2 \equiv 6$};
			\draw [snake=brace] (1.45+3,-7.05) -- (1.45+3,-7.45);
			\draw [snake=brace] (1.45+3,-7.55) -- (1.45+3,-7.95);
			\node at (1.55+3,-7.25) [right] {$10$};
			\node at (1.55+3,-7.75) [right] {$3$};
			\draw [snake=brace] (1.95+3,-7.30) -- (1.95+3,-7.70);
			\node at (2.05+3,-7.50) [right] {$10$};
			
			\node at ( 0   +6,-7.0) [right] {$5\cdot 1 \equiv 11$};
			\node at (-0.17+6,-7.5) [right] {$5\cdot a_0 \equiv 5$};
			\node at (-0.32+6,-8.0) [right] {$5\cdot {a_0}^2 \equiv 10$};
			\draw [snake=brace] (1.45+6,-7.05) -- (1.45+6,-7.45);
			\draw [snake=brace] (1.45+6,-7.55) -- (1.45+6,-7.95);
			\node at (1.55+6,-7.25) [right] {$11$};
			\node at (1.55+6,-7.75) [right] {$5$};
			\draw [snake=brace] (1.95+6,-7.30) -- (1.95+6,-7.70);
			\node at (2.05+6,-7.50) [right] {$11$};
			
			\node at ( 0   +9,-7.0) [right] {$5\cdot 1 \equiv 12$};
			\node at (-0.17+9,-7.5) [right] {$5\cdot a_0 \equiv 7$};
			\node at (-0.32+9,-8.0) [right] {$5\cdot {a_0}^2 \equiv 14$};
			\draw [snake=brace] (1.45+9,-7.05) -- (1.45+9,-7.45);
			\draw [snake=brace] (1.45+9,-7.55) -- (1.45+9,-7.95);
			\node at (1.55+9,-7.25) [right] {$12$};
			\node at (1.55+9,-7.75) [right] {$7$};
			\draw [snake=brace] (1.95+9,-7.30) -- (1.95+9,-7.70);
			\node at (2.05+9,-7.50) [right] {$12$};

			\node at ( 0   +0,-10.5) [right] {$5\cdot 1 \equiv 13$};
			\node at (-0.17+0,-11.0) [right] {$5\cdot a_0 \equiv 9$};
			\node at (-0.32+0,-11.5) [right] {$5\cdot {a_0}^2 \equiv 1$};
			\draw [snake=brace] (1.45+0,-10.55) -- (1.45+0,-10.95);
			\draw [snake=brace] (1.45+0,-11.05) -- (1.45+0,-11.45);
			\node at (1.55+0,-10.75) [right] {$13$};
			\node at (1.55+0,-11.25) [right] {$9$};
			\draw [snake=brace] (1.95+0,-10.80) -- (1.95+0,-11.20);
			\node at (2.05+0,-11.00) [right] {$13$};
			
			\node at ( 0   +3,-10.5) [right] {$5\cdot 1 \equiv 14$};
			\node at (-0.17+3,-11.0) [right] {$5\cdot a_0 \equiv 11$};
			\node at (-0.32+3,-11.5) [right] {$5\cdot {a_0}^2 \equiv 5$};
			\draw [snake=brace] (1.45+3,-10.55) -- (1.45+3,-10.95);
			\draw [snake=brace] (1.45+3,-11.05) -- (1.45+3,-11.45);
			\node at (1.55+3,-10.75) [right] {$14$};
			\node at (1.55+3,-11.25) [right] {$11$};
			\draw [snake=brace] (1.95+3,-10.80) -- (1.95+3,-11.20);
			\node at (2.05+3,-11.00) [right] {$14$};
			
			\node at ( 0   +6,-10.5) [right] {$5\cdot 1 \equiv 15$};
			\node at (-0.17+6,-11.0) [right] {$5\cdot a_0 \equiv 13$};
			\node at (-0.32+6,-11.5) [right] {$5\cdot {a_0}^2 \equiv 9$};
			\draw [snake=brace] (1.45+6,-10.55) -- (1.45+6,-10.95);
			\draw [snake=brace] (1.45+6,-11.05) -- (1.45+6,-11.45);
			\node at (1.55+6,-10.75) [right] {$15$};
			\node at (1.55+6,-11.25) [right] {$13$};
			\draw [snake=brace] (1.95+6,-10.80) -- (1.95+6,-11.20);
			\node at (2.05+6,-11.00) [right] {$15$};
			
			\node at ( 0   +9,-10.5) [right] {$5\cdot 1 \equiv 16$};
			\node at (-0.17+9,-11.0) [right] {$5\cdot a_0 \equiv 15$};
			\node at (-0.32+9,-11.5) [right] {$5\cdot {a_0}^2 \equiv 13$};
			\draw [snake=brace] (1.45+9,-10.55) -- (1.45+9,-10.95);
			\draw [snake=brace] (1.45+9,-11.05) -- (1.45+9,-11.45);
			\node at (1.55+9,-10.75) [right] {$16$};
			\node at (1.55+9,-11.25) [right] {$15$};
			\draw [snake=brace] (1.95+9,-10.80) -- (1.95+9,-11.20);
			\node at (2.05+9,-11.00) [right] {$16$};
			
		\end{tikzpicture}
	\end{center}
	\caption{\bf{Multiplication Results with $\mathbf{a_0=2}$ Modulo 17}\label{TableModulo17}}
\end{PageFigure}

\Section{A Two Clause\\ Walkthrough}
In order to help understand the algorithm, this walkthrough will start at the beginning of execution and proceed through as many steps as possible.  The initial problem will be given as the equation:

\defeq{TwoClauseExampleEquation}
(x_0) \land (\overline{x_0} \lor x_1)
\eeq

One of the first things to do is to determine the modulus.  Using a prime greater than $(2n)^2$ should work well, and it's obvious that there are 2 clauses and 2 variables.  17 should suffice, since $17 > (2\cdot 2)^2$.

Next, a value should be given to $x$, which will be used to calculate the clauses.  There doesn't seem to be any particularly good way to pick, other than using a number greater than one.  Three will be used for this example.

Next, the values for the clauses can be calculated.  Recalling Equation \ref{XmEquation} on page \pageref{XmEquation}:

\defeq{XmEquation2}
f(x_m)=\left(\prod_{k=0}^{m-1}(1+x^{2^k})\right)x^{2^m}\left(\prod_{k=m+1}^{n}(1+x^{2^k})\right)
\eeq

So, 

\begin{align*}
f(x_0) &= \left(1+x^{2^1}\right)x^{2^0} \\
&= \left(1+x^2\right) x^1 \\
&= (1+3^2)3^1 \\
&= (10)3 = 30 \equiv 13 \mod 17
\end{align*}

The second, trickier equation comes from the same subsection.  The algorithm to use is in Figure \ref{NegatedClauseAlgorithm} on page \pageref{NegatedClauseAlgorithm}.  This is modified from Figure \ref{ClauseAlgorithm} on page \pageref{ClauseAlgorithm}.  Here,

\begin{align}
g(\overline{x_0}) = () = 1
\end{align}

There is nothing to multiply, so by convention this product will be set equal to one.  As for $x_1$:

\begin{align*}
g(x_1) &= \left(1+x^{2^0}\right)x^{2^1} \\
&= \left(1+x^1\right)x^2 \\
&= (1+3^1)3^2 \\
&= (4)9 = 36 \equiv 2 \mod 17
\end{align*}

Now, proceeding through the algorithm starting at h=0, that results in the case $h=\overline{a_i}$.  So the result becomes $g(\overline{x_0})$, or 1.  Now, h is incremented, so h=1.  This results in the case $h=a_i$.  So $g(x_1)$ is added to the result, giving $1 + 2 = 3$.  This finishes the algorithm with the result $f(\overline{x_0} \lor x_1) \equiv 3 \mod 17$.

At some time, the equations should be set up so that Boolean algebra can eventually lead to an answer.  Now is a good time, so the various results for $a_0=2$ and $c_0$ are shown in Figure \ref{TableModulo17} on the following page.  The goal here is to eliminate the diagonal by taking combinations of equations, as explored in the previous section.  Observe what happens if another set of equations is created, with $a_1 = 3$ and $c_1 = 1$:

\begin{align*}
c_1 \cdot 1 &\equiv 1 \cdot 1 &\equiv 1 \\
c_1 \cdot a_1 &\equiv 1 \cdot 3 &\equiv 3 \\
c_1 \cdot {a_1}^2 &\equiv 1 \cdot 3^2 &\equiv 9
\end{align*}

Proceeding as in Figure \ref{MultiplicationDifferences} on page \pageref{MultiplicationDifferences}, the first differences of this equation are $3-1=2$ and $9-3=6$.  The second difference is $6-2=4$.  Recalling that the second differences should add up to the modulus (17), a second difference of $17-4=13$ is required.  From Figure \ref{TableModulo17}, it can be seen that setting $c_0=13$ in the first set of equations gives a second difference of 13.  Thus, these two equations can now be combined succesfully.  To see this, simply use the original values of $a$ and $c$ in each equation and add them together.  For example using $a_0=2$ and $c_0=13$ for the first equation, and $a_1=3$ and $c_0=1$ for the second equation, this gives results of combining equations:

\begin{align*}
c_0 + c_1 &\equiv 13 + 1 &\equiv 14 \\
c_0(a_0) + c_1(a_1) &\equiv 9 + 3 &\equiv 12 \\
c_0({a_0}^2) + c_1({a_1}^2) &\equiv 1 + 9 &\equiv 10
\end{align*}

Now corresponding addition equations can be constructed by observing:

\begin{align*}
14 \Longleftrightarrow 14 + 0(-2) \\
12 \Longleftrightarrow 14 + 1(-2) \\
10 \Longleftrightarrow 14 + 2(-2)
\end{align*}

A similar examination can be done to produce two more addition equations.  Proceeding by increasing the previous value of $a$ for each successive multiplication equation, recall that the last multiplication equation used $a_1=3$.  So pick $a_2=4$.  Use $c_2=1$ to start.  This gives:

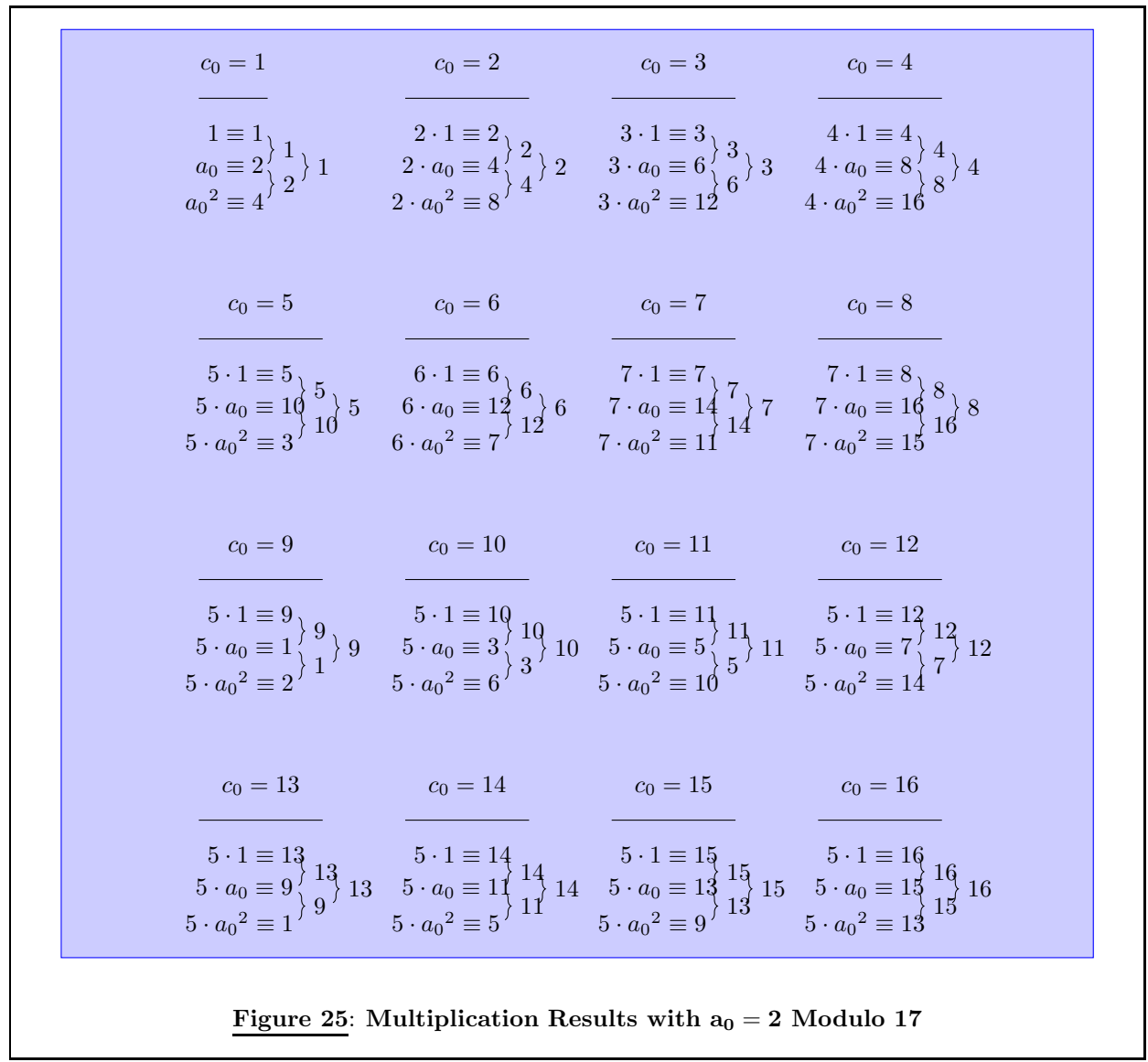
\begin{align*}
c_2 \cdot 1 &\equiv 1 \cdot 1 &\equiv 1 \\
c_2 \cdot a_2 &\equiv 1 \cdot 4 &\equiv 4 \\
c_2 \cdot {a_2}^2 &\equiv 1 \cdot 4^2 &\equiv 16
\end{align*}

This gives first differences of $4-1=3$ and $16-4=12$, and a second difference of $12-3=9$.  So look for another second difference that sums 9 to a total of 17.  $17-9=8$.  Consulting Figure \ref{TableModulo17} again, it can be seen that using $c_0=8$ gives a second difference of 8.  To check this and set up the values for the addition equations, use $a_0=2$, $c_0=8$, $a_2=4$, and $c_2=1$:

\begin{align*}
c_0 + c_2 &\equiv 8 + 1 &\equiv 9 \\
c_0(a_0) + c_2(a_2) &\equiv 16 + 4 &\equiv 3 \\
c_0({a_0}^2) + c_2({a_2}^2) &\equiv 15 + 16 &\equiv 14 \equiv -3
\end{align*}

Once again, observe that there is a difference of successive results by -6.  For example $9 + (-6) = 3$, and $3 + (-6) = -3$.  So the second set of addition equations can be set up:

\begin{align*}
9 \Longleftrightarrow 9 + 0(-6) \\
3 \Longleftrightarrow 9 + 1(-6) \\
-3 \Longleftrightarrow 9 + 2(-6)
\end{align*}

A third set of addition equations can now be set up, which will conclude the setup.  Continuing successively, pick $a_3 = 5$.  With $c_3 = 1$, this gives:

\begin{align*}
c_3 \cdot 1 &\equiv 1 \cdot 1 &\equiv 1 \\
c_3 \cdot a_3 &\equiv 1 \cdot 5 &\equiv 5 \\
c_3 \cdot {a_2}^3 &\equiv 1 \cdot 5^2 &\equiv 8
\end{align*}

It gives first differences of $5-1=4$ and $8-5=3$, and a second difference of $3-4 \equiv 16$.  $16 + 1 = 17$, so a second difference of one is required for another set of equations.  Once again using Figure \ref{TableModulo17}, $c_0=1$ gives this equation for $a_0=2$.  Thus:

\begin{align*}
c_0 + c_3 &\equiv 8 + 1 &\equiv 2 \\
c_0(a_0) + c_3(a_3) &\equiv 2 + 5 &\equiv 7 \\
c_0({a_0}^2) + c_3({a_3}^2) &\equiv 4 + 8 &\equiv 12
\end{align*}

For this last set of equations we observe:

\begin{align*}
2 \Longleftrightarrow 2 + 0(5) \\
7 \Longleftrightarrow 2 + 1(5) \\
12 \Longleftrightarrow 2 + 2(5)
\end{align*}

This setup gives the information needed to set up a system of three equations in three unknowns.  The next step involves preparing the clause polynomials for addition and multiplication.  Once they are prepared, addition and multiplication can be used to set up the linear algebra system.

First off, the function of ones should be calculated.  This is easy, recall Equation \ref{OnesEquation} from page \pageref{OnesEquation}:

\defeq{OnesEquation2}
f(1) = \prod_{k=0}^{V-1}{(1+x^{2^k})}
\eeq

Here the function of ones is:
\begin{align}
f(1) &= \prod_{k=0}^{V-1}{(1+x^{2^k})} \\
&=(1+x^{2^0})(1+x^{2^1}) \\
&=(1+x^1)(1+x^2) \\
&=(1+3)(1+9) \\
&=(4)(10) \equiv 6
\end{align}

Now that the function of ones is calculated, the multiplication equations can be calculated.  

{\color{red}to be completed later...}

\pagebreak

\Section{Finishing The Two Clause Example}
An important observation that is critical to the entire algorithm can be made at this point.  Originally, the gist of the algorithm was to create 3 equations in 3 unknowns.  Unfortunately, the equations cannot be independent of one another, since we can only vary 2 parameters of the 3 equations;  Only the value for $d$, which corresponds with no clauses satisfied, and the corresponding increment can vary.  Since only these two parameters can vary, we must somehow satisfy the 3 unknowns with only two equations.

This is where the observation occurs.  The three unknowns aren't completely independent.  \textit{ One of the 3 unknowns depends upon the other 2.}

In the case of addition, the three unknowns must add up to the function of ones.  This is because all three unknowns, taken together, completely occupy all of the $x$ coefficients, which is exactly what the function of ones represents.  Similarly, in the case of multiplication, the off-diagonals must add up to the function of ones sqaured, minus the diagonal, which is the function of ones.

Let's refine our model.  First, we only need to come up with two equations.  Back around page \pageref{TwoClauseEquation}, and in Figure \ref{TwoClauseEquation}, we came up with 3 variables; $b_0$, $b_1$, and $b_2$.  $b_0$ represents 0 clauses satisfied, $b_1$ represents one clause satisfied, and $b_2$ represents 2 clauses satisfied.  We just realized that these three variables contain one dependent variable among them.  Let's let $b_2$ be the dependent variable, which makes the equation

\defeq{b2transformed}
4 b_0 + 2 b_1 + 0 b_2 = 4 b_0 + 2 b_1 + 0 (f(1)-b_0-b_1)
\eeq

Thus, we can write our old equation in 3 dependent variables as a new equation in 2 independent variables.  So now we only need 2 equations, and we have them.  We can now solve the system and determine the values associated with no clauses satisfied, one clause satisfied, and two clauses satisfied $(f(1)-b_0-b_1)$.  This is just linear algebra, but care must be taken to ensure that only multiplications are performed instead of division, since we are working inside a finite field.

\pagebreak

\Section{Algorithm Finish (Basics)}
In all cases, the algorithm finishes the first portion with a value for $n$ clauses satisfied.  If this value is nonzero, we can conclude that the current Boolean equation can be satisfied.  Otherwise, there is only a 1-in-$p_2$ chance that the current Boolean equation can be satisfied (where $p_2$ is a probability picked ahead of time that will be discussed in greater detail shortly).  But this doesn't return a certificate; that is, an assignment of variables that satisfies the equation.  If we think that the equation can be satisfied, we should return a certificate.

To return a certificate, multiple occurences of the basic algorithm are run.  This proceeds as follows.  First, we assume that the equation can be satisfied; otherwise, simply return that it can't be satisfied and we are done.  So the next step is to take the first value in the original equation and pick a value for it.  We'll pick true, although we could pick false.  Now we rewrite the original Boolean equation with this new value set as true (which is fairly common knowledge, and may be explained in a future version of this paper).  Then we determine if the new equation can be satisfied.  If it can, we proceed to repeat this process with more variables until a certificate is produced.  If the equation can't be satisfied with the variable set as true, we try a new equation with the variable set as false.  If this works, we again proceed on with more variables.  Now there is a slight chance that neither equation seems to be satisfied.  If this is the case, we can try again with another prime.  What's important here is that there is no need to backtrack.

This is a probabilistic algorithm with bounded error.  At the start of the algorithm, we must know the total probability of error, which we will call $p_3$.  Then we can calculate the maximum probability of error for each iteration, which is $p_2$.  Then we can determine how many and what types of primes to use to give us $p_2$ error at each step.  All of this will be discussed after the general $n$ clause algorithm.  For now we remark that we can continue to examine a particular variable assignment only for so long, and then we can simply conclude that we've exceeded the error probability and conclude that we couldn't deduce a certificate; only satisfiability.

The major point to take away here is that an $n$ clause Boolean equation with $V$ variables will eventually require the algorithm for more clauses.  Specifically, this is the $n + 2 V$ clause algorithm.

\pagebreak

\Section{The $n$ Clause Algorithm}
Knowing that the $n$ clause Boolean equation will eventually require the $n + 2 V$ clause algorithm, it's best to anticipate this ahead of time.

The actual general case algorithm works by using many versions of the 2 clause algorithm.  We can begin to see how this can occur by observing that the 2 clause algorithm returns a value for all clauses satisfied, which can be used as one of the four initial values in a second 2 clause algorithm.  The second 2 clause algorithm has the same general requirements as any 2 clause algorithm.  It requires a prime, which we have.  It requires the information for 2 clauses; two numbers for addition, and two numbers for multiplication.  We can get this by using 4 versions of the 2 clause algorithm, which prepare another iteration of the 2 clause algorithm.  Then we proceed again.  In this fashion, a tree of 2 clause algorithms can be built up to solve an $n$ clause system.

\Subsection{Semi-optimized Version}
The version presented in this paper has only minor optimizations to enhance it; however, there will be much room left for further improvement.

So far, the best way to optimize seems to be to focus on the main portion of the algorithm, which is the 2 clause algorithm.  If we briefly analyze the occurences of this, we should know that it combines two clauses into one output.  Each clause needs a portion defined for addition, and another for multiplication.  So for every 2 clauses in, there are 4 total inputs.  This then leads to a single output.  So for $n$ clauses, at every level the number of clauses is reduced by half.  Similarly, the number of inputs is reduced to a fourth.  We can designate the number of such levels as $l$.  Now for $n=2^l$ clauses there are $l$ levels to the tree.  Similarly, there are $4^l$ inputs needed for all of the leaves.  We note that the $4^l$ inputs is equal to the $2^l$ clauses squared: $4^l = (2^l)(2^l)$, just as the inputs are the number of clauses squared.  So we can conclude that there are roughly $n^2$ total instances of the 2 clause algorithm that occur for the $n$ clause algorithm.  We also know that the $n$ clause algorithm will be run $O(V)$ times in the case of a satisfied Boolean equation in order to determine a certificate (a set of satisfying variables).  So we can also conclude that the algorithm will run the 2 clause algorithm no more than $O(V)(n+V)^2$ times.

Knowing that the current center of attention for the $n$ clause algorithm is the repetition of the 2 clause algorithm, we can optimize the performance of the 2 clause algorithm by making some precomputations.  This is because we can use the same values repeatedly for the 2 clause algorithm, with the only exceptions (outside of intermediate computations) being the 4 inputs and single output.

Knowing that the main algorithm will call the two clause algoithm no more than $O(V)(n+V)^2$ times, we can set all of the primes we use to be approximately this value.  Thus we set our primes that we use to be $\Theta(V(n+V)^2)$.  All calculations can now be performed knowing the value of the prime ahead of time, and this will also help to ensure that the mulitplication inputs are rarely ever zero.  In the case that they are, the algorithm will have to perform some addition steps to try to ensure that they become nonzero.

Returning to the precalculations, we can set up the equations to proceed as quickly as possbile.  Calculate the appropriate $a_k$'s and $c_k$'s so that all two clause algorithms run with the same equations.  In fact, all of the constants can be precalculated individually in $O(V)(n+V)^2$ time by simply cycling through every possible natural (plus zero) up to the prime $p$ and picking the appropriate value.

So the precalculations all take $O((V)(n+V)^2)$ time.  Note that the two clause algorithms, at this point, should take constant time.

The algorithm is almost complete.  Only one important piece remains; the problem that arises when a zero is given as an input for multiplication.  There is one good possibility that we can make use of.  \textit{We use extra variables in our calculations.}  To do this, we simply perform clause calculations with double the number of variables in the equation (A different value could be used, but this seems to be a fairly useful amount).  Now, when a zero comes up as an input for multiplication, we can simply add in a new equation with the new variables.  If the original equation is satisfiable, this should help to change the multiplication input.  Otherwise, we will conclude that the equation is questionably unsatisfiable  (It is unsatisfiable to within the error probability $1/p$).

\Section{Runtime Analysis /\\ Correctness}
As mentioned and explained in the previous section, most components take $O(V(n+V)^2)$ time.  The error correction in the case of a zero input correction is seperate from the main algorithm, and obviously can be done in $O(V(n+V)^2)$ time.  So for a single prime, the runtime is $O(V(n+V)^2)$.

This is a nonrandomised, deterministic algorithm with bounded error.  Each prime used gives an individual error bound of $1 / \Theta(V(n+V)^2)$.  Together, $P$ primes give approximately an error bound not exceeding probability $1 / \Theta(V(n+V)^2)^P$.  So the algorithm runs in time $O(P\cdot V(n+V)^2)$ with mistaking satisfiable Boolean expressions as unsatisfiable with an approximate probablity $1 / \Theta(V(n+V)^2)^P$.
 
\pagebreak 

\Section{Conclusion}
I hope that this project presents sufficient evidence that P=NP.  I've written the last few pages rather hastily, but hope to improve things soon.  I'm starting on writing the code for this project, so that everything can be put under better scrutiny.

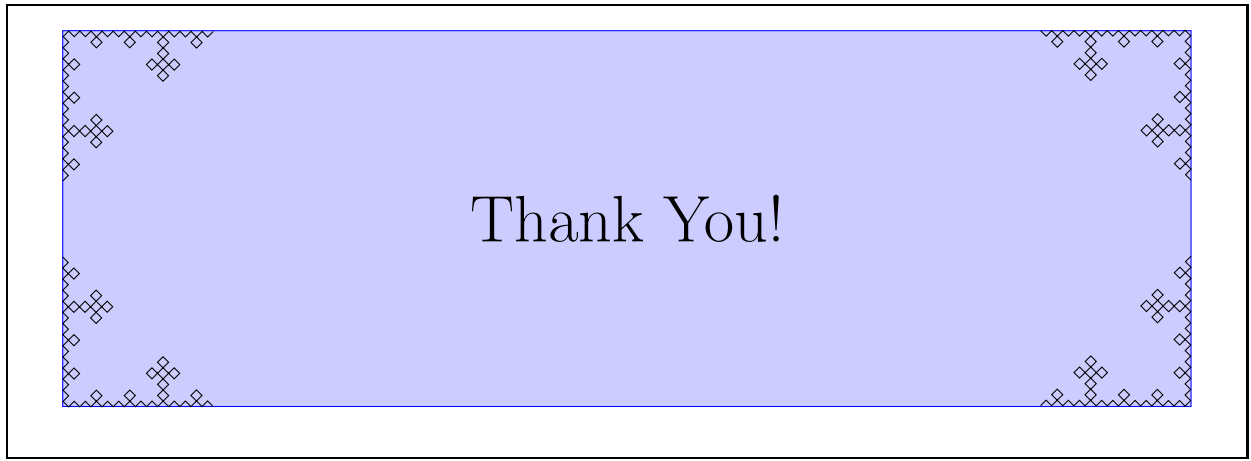
\begin{table*}
\begin{framed}
	\begin{center}
		\begin{tikzpicture} 
			\filldraw[fill=blue!20, draw=blue] (0,0) rectangle (15,-5);
			\draw [decoration=Koch curve type 1] decorate{ decorate{ decorate{ (0,-2) -- (2,0) }}};
			\draw [decoration=Koch curve type 1] decorate{ decorate{ decorate{ (13,0) -- (15,-2) }}};
			\draw [decoration=Koch curve type 1] decorate{ decorate{ decorate{ (15,-3) -- (13,-5) }}};
			\draw [decoration=Koch curve type 1] decorate{ decorate{ decorate{ (2,-5) -- (0,-3) }}};
			
			\node at (7.5,-2.5) {\Huge Thank You!};
			
		\end{tikzpicture}
	\end{center}
\end{framed}
\end{table*}

\pagebreak

\Section{Acknowledgements}
I would like to thank God and my family for helping to provide me with this wonderful opportunity.  I would also like to thank Javier Humberto Ospina Holguin for showing me an interesting new world and encouraging me to get involved.  I would like to also thank the creator of Bricks; Andreas Rottler, for providing a great game that helped get me excited about problems like this, as well as a way to meet interesting people such as Javier.

I would also like to thank the people who helped to save my life when I was in danger;  the Mexican/American family in Putla, Holy Spirit Hospital (which also helped me with my schizophrenia), and some people from Harrisburg, Pennsylvania.

I'd like to thank Timothy Wahls of Dickinson College (formerly of Penn State Harrisburg) for first introducing me to the problem and getting me excited.

I'd also like to thank my friend Holly Dudash for being with me during tough times and helping me through.

I'd like to thank Michael Sipser of MIT and my friend Ruben Spaans for analyzing my ideas and their invaluable suggestions.

I'd like to thank the many great organizations, communities, and schools that helped me learn and fostered my intellectual growth including SOS Mathematics (online), Stackexchange - Mathoverflow.com, stackoverflow.com, and cstheory.stackexchange.com.  Particularly Ryan O' Donnel, Arturo Magadin, Carl Brannen, and Qiaochu Yuan.  Also Penn State University, and in particular Penn State Harrisburg.  Especially Drs. Null, Bui, Walker, and Wagner.  Also from other various Universities Jacques Carrette, George Frederick Viamontes, Will Jagy, and Leonid Levin.

I'd like to thank everyone that has worked on solving the problem(s) of schizophrenia, and I hope that this paper (although perhaps indirectly) will help with more research.

There are many more people that I regretably haven't mentioned, but I hold them in high esteem and send out my thanks to.  I'm just very thankful that I have a chance to be a part of a project that will hopefully do lots of good things.

\pagebreak
\appendix
\Section{Clause Polynomial\\ Technicalities\label{FormulaAppendix}}

\pagebreak
\Section{Function of Ones\label{OnesFunction}}

\pagebreak

\pagebreak
\pagebreak




\end{document}